\documentstyle[aps,prd,preprint,epsf]{revtex}

\preprint{DAMTP--2001-65} 
\date{\today}
\tighten

\newcommand{\ea}{\textit{et al.}}
\newcommand{\AH}{Abelian-Higgs}

\begin{document}

\title{On the Evolution of Abelian-Higgs String Networks}

\author{J.N.~Moore${}^{1}$\thanks{
Electronic address: J.N.Moore\,@\,damtp.cam.ac.uk},
E.P.S.~Shellard${}^{1}$\thanks{
Electronic address: E.P.S.Shellard\,@\,damtp.cam.ac.uk}
and C.J.A.P.~Martins${}^{1,2}$\thanks{
Electronic address: C.J.A.P.Martins\,@\,damtp.cam.ac.uk}}

\address{${}^1$ Department of Applied Mathematics and Theoretical Physics\\
Centre for Mathematical Sciences, University of Cambridge\\
Wilberforce Road, Cambridge CB3 0WA, U.K.}

\address{${}^2$ Centro de Astrof\'{\i}sica, Universidade do Porto\\
Rua das Estrelas s/n, 4150-762 Porto, Portugal}

\maketitle
\begin{abstract}
{
We study the evolution of Abelian-Higgs string networks in large-scale
numerical simulations in both a static and expanding background. 
We measure the properties of the network by tracing the motion of the
string cores, for the first time estimating the rms velocity of the 
strings and the 
invariant string length, that is, the true network energy density.
These results are compared with a 
velocity-dependent one scale model for cosmic string network
evolution.  This incorporates the contributions of loop production,
massive radiation and friction to the energy loss processes that are
required for scaling evolution. We use this analysis as a basis for 
discussing the relative importance
of these mechanisms for the evolution of the network.
We find that the loop distribution
statistics in the simulations are consistent with the long-time
scaling of the network being dominated by loop production.  
Making justifiable extrapolations to cosmological scales, these
results appear to be consistent with the standard 
picture of local string network evolution in which loop production
and gravitational radiation are the dominant decay mechanisms.
}
\end{abstract}
\pacs{PACS number(s): 98.80.Cq,11.27.+d}

\section{Introduction}
\label{secintro}

Vortex-string networks in three dimensions are important in a variety of
contexts, whether in condensed matter physics or cosmology (for reviews 
see \cite{Vilenkin1994cossot,Hindmarsh1995re,Geyer1995fiettc}).
If we are to obtain a
quantitative description of these networks, then we must develop an
adequate understanding of their `scaling' evolution as well as the decay
mechanisms which maintain it. The \AH{} model, a
relativistic version of the Ginsburg-Landau theory of
superconductors, provides a convenient testbed for developing
detailed models for this evolution. On the one hand, the relatively
simple field theory can be studied directly in three-dimensional
simulations.  On the other, a straightforward reduction to a
one-dimensional effective theory---the Nambu action---can also be
studied numerically, though over a much wider dynamic range.

In a cosmological context, a rather simple `one-scale' model of
string evolution has emerged \cite{Kibble1985hp,Martins1996jp}
which appears to successfully describe the large-scale features of
an evolving string network \cite{Bennett1990yp,Allen1990tv}, though
with subtleties remaining on smaller scales.  In this simple model,
after an initial transient period,
the average number of long strings in a horizon volume remains fixed
as it expands, a rapid dilution made possible through re-connections
resulting in loop production. The loops, in this standard picture,
oscillate relativistically and decay through gravitational
radiation.  The subtlety here concerns the length scale at which
small loop creation occurs and whether this is inherently scale-invariant
\cite{Austin1993rg,Martins2000mspric}. Even if this were not so, additional gravitational radiation back-reaction effects should
act on the long string network, eliminating small
wavelength modes and thus setting a minimum loop creation
size which is scale-invariant\cite{Bennett1990yp,Austin1993rg}. 
In any case, high resolution Nambu string simulations have found 
that large-scale
network properties are relatively insensitive to this loop creation
scale, below a minimum loop size \cite{Bennett1990yp,Allen1990tv}. 

This standard picture for network evolution has been
questioned on the basis of \AH{} field theory
simulations \cite{Vincent1998cx}, as well as the correspondence
between the Nambu simulations and global field theory simulations
\cite{Yamaguchi1999yp,Yamaguchi1999dy}. In the first case \cite{Vincent1998cx},
the authors suggested that the
primary energy loss mechanism by long strings is direct massive
radiation, rather than loop creation
(we denote this as the `scale-invariant massive radiation' scenario). 
This is contrary to previous 
expectations that the presence of a large mass threshold will
strongly suppress massive particle production for long wavelength
oscillatory string modes, that is, those much larger than the string
width \cite{Srednicki1987xg,Moore1998gp,Olum1999sg,Moore2000a}.
Evidence presented in support of this includes Smith-Vilenkin
simulations in which scaling is observed only if strings are
produced at the smallest available scales, the profusion of loops at
resolution scale in other Nambu codes, field theory network
simulations in which the loop density is observed to be low, and the
study of large amplitude oscillations of a single string.

The primary purpose of this paper is to study an evolving string
network and its decay mechanisms.  This involves extensive
large-scale numerical simulations of \AH{} string networks in flat
and expanding backgrounds.  Simulations in an expanding universe, in
particular, are severely limited in dynamic range so we also explore
a new ``fat string'' algorithm appropriate for gauged strings, which
keeps the string width constant in comoving coordinates.  
For comparison with a theory
that is known to be strongly radiating, we simulated global strings
in an expanding background (studied previously by Yamaguchi and
collaborators \cite{Yamaguchi1999yp,Yamaguchi1999dy}).  Having
characterised the initial scaling properties of an \AH{} network, we
endeavour to model the dominant decay mechanisms using the
velocity-dependent one-scale (VOS) model.  In order to achieve this,
we have developed a new suite of diagnostic tools to estimate both the
(static) string length and the average rms velocity of the network.
The latter is challenging and has not been attempted in previous
work but it is important, first, if we are to estimate the true (invariant)
string length and, secondly, because it can be used to distinguish between
different network decay mechanisms.  We find some reasonable
best-fit parameters for the VOS model which are self-consistent
across all backgrounds (flat or expanding) and, on this basis, 
we are able to eliminate a number of
scenarios.  Finally, we conclude with a qualitative study of loop
production in these networks. We emphasise the care which must be
exercised in trying to extrapolate these field theory simulation
results by many orders of magnitude to cosmological scales.

\section{String network evolution}
\label{secnetwork}

\subsection{Simulation methods}
\label{secmethods}

The work in this paper involves simulations of
systems, in flat or FRW backgrounds, arising from the \AH{}
Lagrangian:
\begin{equation}
  {\mathcal{L}} =  D_{\mu}\phi^*D^{\mu}\phi 
  - \frac{1}{4}F_{\mu\nu}F^{\mu\nu}
  - \frac{\lambda}{4}(|\phi|^2 -\eta^2)^2,  \label{ahlag}
\end{equation}
which yields the corresponding equations of motion,
\begin{eqnarray}
  \label{ahfe1}
  D_\mu D^\mu \phi
  &=&  - \frac{\lambda\phi}{2}(|\phi|^2 -\eta^2), \\ 
  \label{ahfe2}
  J^\mu \equiv \partial_\nu F^{\nu\mu}
  &=&    2e {\rm Im} (\phi^* D^\mu \phi).
\end{eqnarray}
Allowing for the rescaling of coordinates and the Higgs field, the only 
free parameter in the model is the ratio $\beta \equiv \lambda/2e^2$.
The critial value $\beta =1$ is the well-known Bogomolyni limit in which 
the scalar and vector forces between like vortices cancel.

Following Moriarty \ea{}~\cite{Moriarty1988fx}, it has become
standard to simulate the dynamics of the abelian-Higgs model
using an approach derived from lattice gauge theory. 
The Lagrangian density is transformed to a Hamiltonian
one and discretised on a lattice of spacing $d$, in the
gauge $A_0=0$. With this formalism the Higgs field variables are associated with
the vertices of the simulation lattice, while the gauge fields are
associated with the links.  To evolve the initial configurations forward in 
time we use a
leap-frog method, using centered derivatives for the time evolution.
The configuration space and momentum space variables are
interleaved, so that for a time-step of $ht$, they are offset by a
time $ht/2$ with respect to each other.  We imposed periodic boundary 
conditions on the string networks.
We describe the specific numerical algorithms 
in much greater detail elsewhere \cite{Moore2000a}, while here we 
focus on important 
modifications required for studying string networks in an expanding 
universe.

In a Friedmann-Robertson-Walker background with scalefactor $a(t)$, the 
evolution equations (\ref{ahfe1}-\ref{ahfe2})
take the following form
\begin{eqnarray}
  \phi''+2\frac{a'}{a}\phi' - D_i D_i \phi & = -\beta a^2 \phi(|\phi|^2-1)   \\
  \partial_0 F_{0i}
  & =  2a^2{\rm Im}(\phi^*D_i\phi)+\partial_jF_{ji}.
  \label{ahfrwfe}
\end{eqnarray}
Here, we have chosen comoving coordinates ${\bf x}$ (${\bf r} = a\, {\bf x}$)
and conformal time $\tau$ ($dt = a\, d\tau$).  The Hubble expansion 
has two separate effects, damping oscillations
in the Higgs field and increasing the effective field masses in direct
proportion to $a$. No damping applies to the Maxwell term.
The equations (\ref{ahfrwfe}) have previously been 
simulated in two
dimensions using the Lorentz gauge  to study vortex formation~\cite{Ye1990na}.
Here, we incorporate the effects of the expanding universe into the 
hamiltonian equations. 

The difficulty of simulating  (\ref{ahfrwfe}) with vortex-strings in 
comoving coordinates is that there is an inevitable loss of 
dynamic range caused by the increase in the effective
field masses as the universe expands.  This can be avoided in a
manner described for global fields in ref.~\cite{Press1989dynedw} 
for the evolution of global domain walls in two dimensions (see 
also~\cite{Ryden1990evondw} and 3D global field theory implementations
\cite{Spergel1991ee,Yamaguchi1999yp}).  The approach
entails treating the coefficients in the field equations that depend
on the scale-factor as coefficients in a systematic way. Applied to
(\ref{ahfrwfe}) this introduces new constant coefficients
$r_\phi,s_\phi,r_A,s_A$, with modified evolution
equations 
\begin{eqnarray}
  \phi''+r_\phi\frac{a'}{a}\phi' - D_i D_i \phi 
	&= -\beta a^{s_\phi} \phi(|\phi|^2-1),   \\
  \partial_0 F_{0i}  +r_A\frac{a'}{a}F_{0i}
  &=  2a^{s_A}{\rm Im}(\phi^*D_i\phi)+\partial_jF_{ji}.
\end{eqnarray}
By choosing $s_\phi=s_A=0$, static solutions in flat space remain as
solutions in the expanding universe. Regarding the expansion of the
universe as an adiabatic parameter change to the equations of
motion, $s_\phi$ and $s_A$ may be tuned so that the time behaviour
of selected phenomena matches that in the expanding universe. For
example, in the expanding universe the proper momentum of a moving
straight brane decays as $a^{-1}$ so that the speed $v$ of a
straight string goes as
${v}/{\sqrt{1-v^2}} \propto a^{-2}$.
Matter fields behave as $\langle \phi-\phi_0\rangle \propto a^{-3}$.
The simulations that have been performed used the values
$r_\phi=2,r_A=0$, as inherited from the FRW Lagrangian equations (the 
Hamiltonian equations were written in a manifestly gauge invariant
form).  
Note that with global strings it is clear that $r_\phi=2$ will
reproduce the appropriate evolution and also the decay
of the massless modes~\cite{Spergel1991ee}. These remarks
may not carry over to the gauged string case.
Varying the relation $r_\phi$ and $s_A$ effectively changes the
definition of the conserved current. From comments in
ref.~\cite{Press1989dynedw}, the network simulation results are not
expected to depend sensitively on $r_\phi$ and $r_A$, but this
remains to be investigated in more detail numerically.

We have performed network simulations using periodic cubic lattices
of side length 250 and greater. The principal physical parameter in
each simulation was the initial correlation length.  To establish a
suitable network configuration we took an initial configuration with
a flat initial power spectrum of fluctuations in the Higgs field,
centred around zero, and zero gauge field.  The spectra were cut
off at large momenta. This improves the stability of the evolution
at fixed amplitude, permitting acceleration of the system's
relaxation to a broken symmetry state by choosing a larger baseline
power for the fluctuations. The modes cut out were of high enough
wave number that they would, under exact dissipative evolution,
decay exponentially.

Note that in contrast to other
work \cite{Vincent1998cx,Yamaguchi1999yp,Yamaguchi1999dy}, the
simulated phase transition and relaxed initial conditions were
usually achieved through pure gradient flow (first-order diffusive
evolution).  We describe the modifications required to achieve this 
elsewhere \cite{Moore2000a}. However, we note that creating these
very quiescent initial conditions was a 
fairly costly process numerically, requiring a 
significant proportion of the total simulation time.  
\begin{figure}
\vbox{\centerline{
\epsfxsize=0.8\hsize\epsfbox{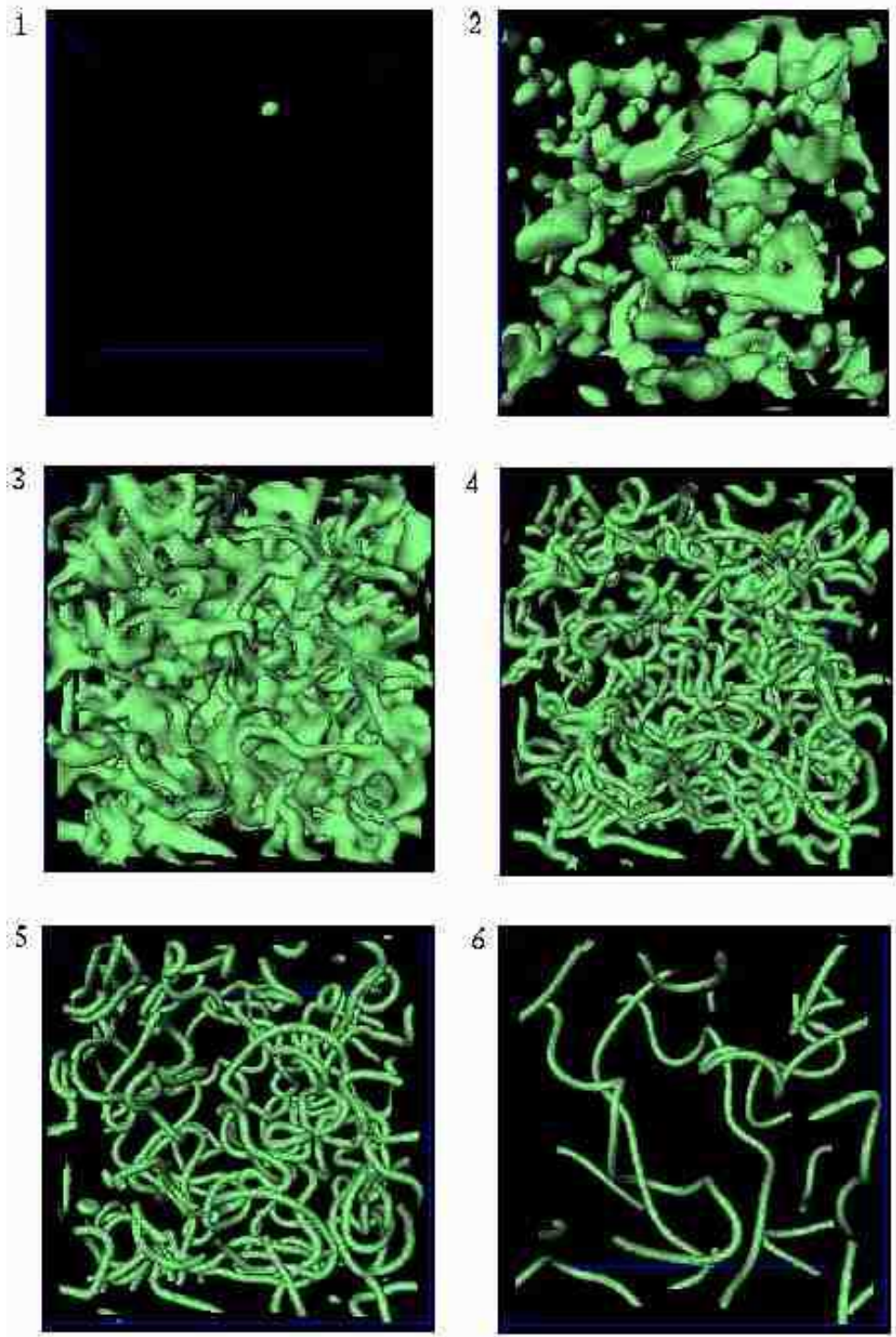}}
\vskip0in}
\caption{String network formation in the \AH{} model using gradient
flow (diffusive) evolution.
Given random initial phases, a symmetry-breaking phase transition
occurs, strings form and then begin to evolve in a scale-invariant
manner (the correlation length is $L\propto
t^{1/2}$). This dissipative evolution is used to create the initial
configuration for a string network with a specified $L$ for subsequent
relativistic evolution.}
\label{figstrings_energy}
\end{figure}

Other feasible methods to set up the initial network 
include the imposition
of a friction term in the second-order evolution equations, or
alternating dissipative evolution with full dynamic evolution.
However, using gradient flow ensured that the density of background
radiation was minimal at a fixed inter-string spacing.  In
particular, it gave greater confidence in our subsequent discussion
when we were able to eliminate friction as a primary network decay
mechanism.
As an aside, we note that the evolution of the network during
gradient flow, as shown in Fig.~\ref{figstrings_energy}, was consistent
with a scaling law in which the correlation length grew as $L\propto
t^{1/2}$ and the velocity decayed as $v\propto t^{-1/2}$.  This
scaling law is well-known from condensed matter physics and also
predicted in this context in the velocity-dependent one-scale
model \cite{Martins1996jp} (see later).

After creating the initial configuration through gradient flow,  
each simulation was evolved using Hamiltonian dynamics
as discussed previously.
The Gauss constraint was satisfied initially as we start from rest,
and was accurately preserved by the equations of motion.  The time-steps 
were chosen to be sufficiently small to ensure the Hamiltonian to be
conserved within 1 per cent over the course of each simulation.
The choice of spatial lattice spacing was constrained by the desire
to meet two conflicting criteria: (i) to simulate a volume which is
orders of magnitude larger than the string width, and (ii) to
accurately represent the continuum theory.  We believe that a physical
lattice spacing of $\Delta x= 0.5$ ($\beta=1$) is close to the maximum 
that is reasonable
given that at larger spacings there is a significant potential
barrier associated with the lattice.  Moreover, for oscillating strings
of length $\approx 15$ we have observed greatly increased radiation
using a larger lattice spacing \cite{Moore2000a}.

There are three important parameters that determine the dynamic
range of the simulations, once the size of the simulation grid and
the initial physical separation of the strings have been determined.
These are the lattice spacing in co-moving coordinates, and the
starting and ending times for the simulations.  These are determined
by specifying the initial string scaling density $\rho t^2/\mu$, and
requiring, at the end of the simulation, that the physical lattice
spacing is 0.5, and that the causal horizon has just propagated to
extend across the simulation box.  Of course, for the `fat string'
algorithm described above, the string remains a fixed size in 
comoving coordinates, so the second constraint is circumvented.

\subsection{String density and velocity analysis}

Here we describe in detail how we analyse the string network
positions, lengths and velocities.  In summary, to characterise the
network configuration at a given point in time we assign positions
of zeros of the Higgs field to lattice plaquettes according to the
winding of the Higgs phase around each plaquette. This allows us to
calculate correlation lengths and loop distribution statistics. When
combined with similar data from a nearby time-step we can also
estimate the velocity of each string segment. As we shall see, the
procedure for calculating the velocities is relatively vulnerable to
numerical normalisation errors due to uncertainty in the precise position of the
string cores.

\begin{figure}
\vbox{\centerline{
\epsfxsize=0.7\hsize\epsfbox{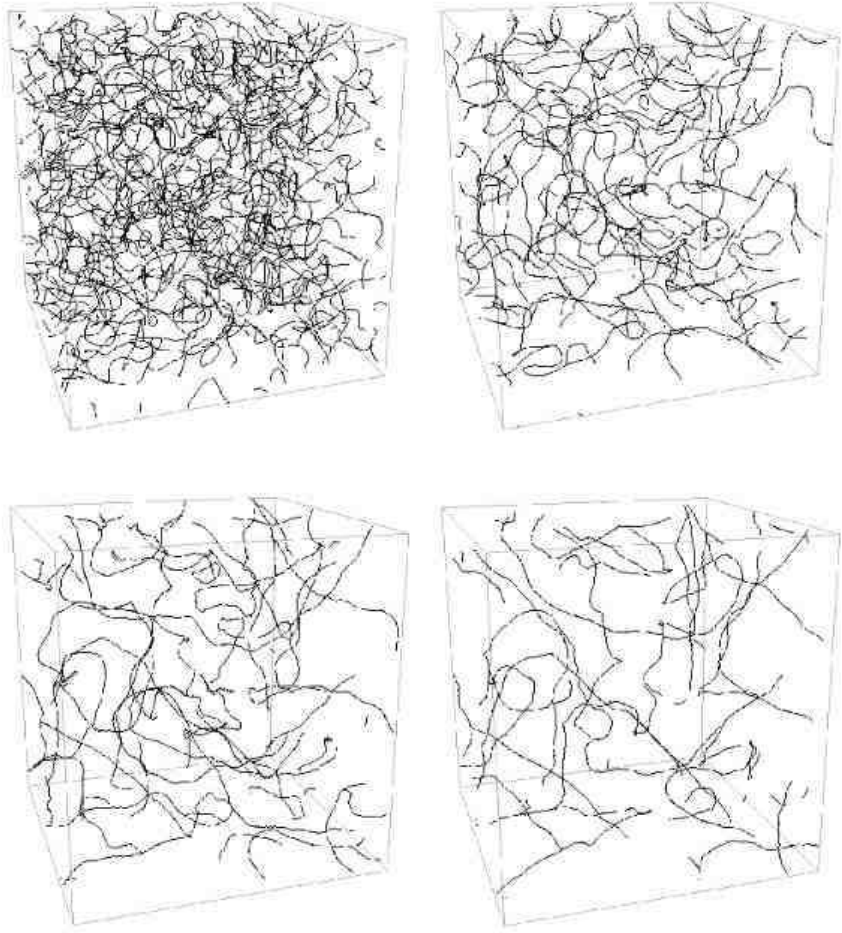}}
\vskip0.2in}
\caption{String network evolution during the radiation era.
Positions for the string cores were found using phase information
and the trajectories connected appropriately.  Note that
lattice discretisation effects have been reduced by smoothing.
The respective times for the advancing evolution are the
conformal times $ t= 24, 32, 40, 48$, by which time the horizon
is comparable to the box size.}
\label{figstrings_core}
\end{figure}

\subsubsection{Static string length estimation}

Positional information for the string was found using the
Vachaspati--Vilenkin algorithms as implemented in the
Allen--Shellard Nambu string network code \cite{Allen1990tv}.
Instead of the random generation of phases, the \AH{} field
theory simulation phases were input and the trajectories of string
zeros approximately reconstructed within the accuracy of the
lattice discretisation.  These trajectories had several string
points assigned for each original lattice spacing, with corners
automatically rounded.  This lattice assignment, however,
necessarily creates straight strings along preferred lattice
directions and wiggly strings along diagonals.  This is not
satisfactory either for estimating the static string length or the
string velocities.  A subsequent period of smoothing or
`point-joining' was undertaken to remove most of these small-scale
artifacts on length scales smaller than the actual string width
$\delta$, while minimising effects above $\delta$.  In practice, it
was not difficult to tune the smoothing to end after rapid
elimination of most of the `wiggliness', before passing over to a slowly
decaying phase.  The evolution of an \AH{} string network
using these positional estimators is shown in 
Fig.~\ref{figstrings_core}.  Note that the strings still exhibit a
small amount of `wiggliness' beyond that inherent in the
pixelization of the rendered image.

Given this information about the string positions and trajectories,
it was a straightforward matter to analyse the total string length
and that of individual loops.  In particular, we could separate the
network into `infinite' loops stretching across the periodic box and
analyse the distribution of small loops.  In terms of the static
configuration, we believe our prescription for estimating the string
length is accurate to of order 10\%.

\subsubsection{String velocity estimation}

Obtaining velocity estimates from this positional information proved
to be challenging, with the overall normalisation remaining the key
uncertainty.  However, the importance of attempting this is obvious, with 
one application being the ability to measure 
the true invariant string length. This
differs from the static string length by an average Lorentz factor
$\bar \gamma \approx (1- v^2)^{-1/2}$, where $v$ here is the rms
velocity of the network weighted over the covariant string length.
The total string energy $E = \mu/L^2$ is clearly related to the
static string energy $E_{\rm s}$ by the relation
\begin{equation}
E = \bar \gamma E_{\rm s} \equiv 
\mu\bar\gamma/L_{\rm s}^2\,,\label{static_energy}
\end{equation}
where $L_{\rm s}$ is the static string correlation length apparently
used in
refs.~\cite{Vincent1998cx,Yamaguchi1999yp,Yamaguchi1999dy}.
Hence, given a typical network velocity of $v\approx 1/\sqrt{2}$, a
static analysis will underestimate the true string energy density
$\rho$ by 40\% and overestimate the correlation length by 20\%.
Furthermore, the static analysis of
refs. \cite{Yamaguchi1999yp,Yamaguchi1999dy} did not measure actual
string trajectories, but rather the proportion of the numerical
volume above a particular threshold potential energy.  By comparing
with a static string profile, an estimate of the string length was
made.  This appears, however, to neglect a further Lorentz factor
since the string cross-section is Lorentz contracted in the
direction transverse to its motion.  Our expectation of a difference
of a factor of two between our density estimates and those of
ref. \cite{Yamaguchi1999yp} is borne out (refer to the global
strings analysis later).

Apart from the overall density normalisation, a static analysis cannot
satisfactorily probe the initial relaxation when the velocity is
rising from zero to its scaling value.  Given the severely
restricted dynamic range possible in field theory simulations, this
initial relaxation is up to 30\% of the total time period available.
Both these reasons, then, compel us to make the 
first attempt to calculate network velocities.

String velocities were estimated by comparing positional information from
two simulations with a small temporal separation $\Delta t$, but
sufficient to ensure that strings have on average moved more than one
lattice spacing $v \Delta t \gtrsim d $. For each individual string 
point $i$ in one simulation, we
undertake an exhaustive search for the nearest two strings points in
the second simulation.  This is achieved with a radix search
implemented in the Nambu code \cite{Allen1990tv} which is of order
$N\log N$, where $N$ is the number of string points.  Having found
the nearest points we connect them with a straight line and find the
perpendicular distance $d_i$ to the original point.  If this
distance is such that there could be no causal relationship we
reject that point (for example, if a small loop has disappeared
during $\Delta t$), otherwise we accept this velocity estimate $v_i
= d/\Delta t$.  We continue iterating over all string points to find
the average root mean square velocity $v$ for the whole string
network.

\begin{figure}
\vbox{\centerline{
\epsfxsize=0.7\hsize\epsfbox{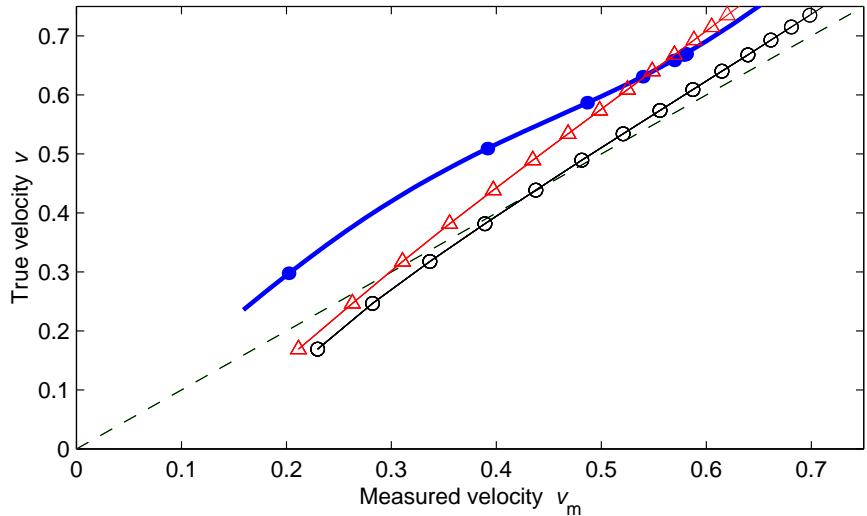}}
\vskip0in}
\caption{Correction to the measured velocity required for
the velocity-position diagnostic to obtain the true velocity
(solid blue line).
This data represents the velocity 
inferred from the growth in the kinetic energy for an initially 
static network (see Fig.~\ref{figv_normalization}). 
Tests of the diagnostic for a distribution of straight string segments
(circles) showed almost direct proportionality, 
whereas idealized small and large rings (red triangles)
also underestimated the true velocity.
}
\label{figv_correction}
\end{figure}

This velocity is a non-trivial statistical estimate which, because
of lattice discretisation effects, only converges in the limit of
measuring many random string positions and orientations.  To test
the reliability of the method we generated field phase information
for a randomly oriented straight string with a given speed $v$ and
showed convergence was adequate for $N\gtrsim 1000$ realizations.  
However, at each
velocity there is a systematic offset between the measured rms
velocity $v_{\rm m}$ and the true rms velocity $v$.  These 
corrections are important, for example, at 
small velocities where jumps between discrete lattice centres tend
to overestimate the velocities, but there are also systematic
effects at high velocity.  A comparison of the inferred velocity from 
an ensemble of moving straight line segments demonstrated the 
efficacy of this algorithm for estimating the true string velocity
in this idealised case (refer to Fig.~\ref{figv_correction}).  This 
was also the case for large circular rings with a typical correlation 
length found in the field theory simulations.  However, for very small 
rings (at about 1/3 the typical correlation length) the diagnostic
underestimated high velocities, thus respresenting a shortcoming for 
highly curved regions on the string network.  The estimated velocity
for an admixture of equal 
numbers of large rings with the correlation length ($L=7.5$) and 
small rings ($L=2.5$) illustrates this effect in Fig.~\ref{figv_correction}.

\begin{figure}
\vbox{\centerline{
\epsfxsize=0.7\hsize\epsfbox{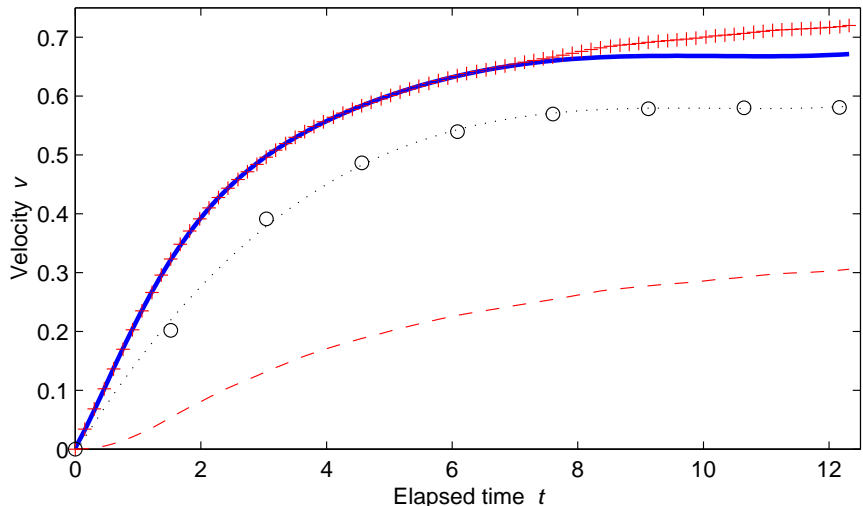}}
\vskip0in}
\caption{The string velocity (solid blue line) inferred from the growth 
in kinetic
energy of a network field theory simulation in flat space
(shown as red '$+$' data points).  The actual 
proportion of kinetic energy relative to the total energy is the red 
dashed line.  The raw velocity diagnostic before normalisation (circular data
points) underestimates the true rms network velocity by about 15\%.}
\label{figv_normalization}
\end{figure}

Given these geometrical effects which are not easily quantifiable in 
a complex string network, we sought an unambiguous 
normalisation for our velocity diagnostic. 
This only proved possible by a direct comparison with the 
overall velocity inferred from the kinetic energy of the full field
theory simulations in flat space.  During the first few time steps as the 
initially static string configuration begins to accelerate and 
move relativistically, the rapid rise
in the kinetic energy is due primarily to string motion.  
Although some radiation will be produced, it is absent in the initial 
conditions and there is insufficient time for this 
to be produced in quantity.  The proportion of the total energy in this
kinetic energy is shown for a large-scale 
flat space simulation in Fig.~\ref{figv_normalization} (the red 
dashed line); the initial correlation length is $L=7.5$.   
Consistent with the rise in the kinetic energy
being due to string motion, initially there is almost no 
change in the overall 
invariant length of the string network.  Plotted as a solid blue line in 
Fig.~\ref{figv_normalization} is the velocity inferred from modelling
the kinetic energy as a change in the string Lorentz factor.  This
should be valid up to about $t\lesssim L = 7.5$, but only a crude upper bound 
thereafter.  By comparing our velocity
estimator which uses positional information from the string cores, we
can hope to normalise it correctly for a general network 
configuration.  The rescaling shown is about 15\%
higher than the straight string normalisation discussed above. It yields
results consistent with the expected $v\approx 1/\sqrt{2}$ for 
flat space networks and we believe to be accurate within a 10\% range.
However, we cannot exclude at this stage the possibility 
that the velocities quoted in what follows are slightly 
overestimated (or, less likely, underestimated).
There may be unexplored physical mechanisms which can systematically 
reduce velocities on the small-scales probed by the 
present simulations.

\subsection{Flat space networks}

\begin{figure}
\vbox{\centerline{
\epsfxsize=0.7\hsize\epsfbox{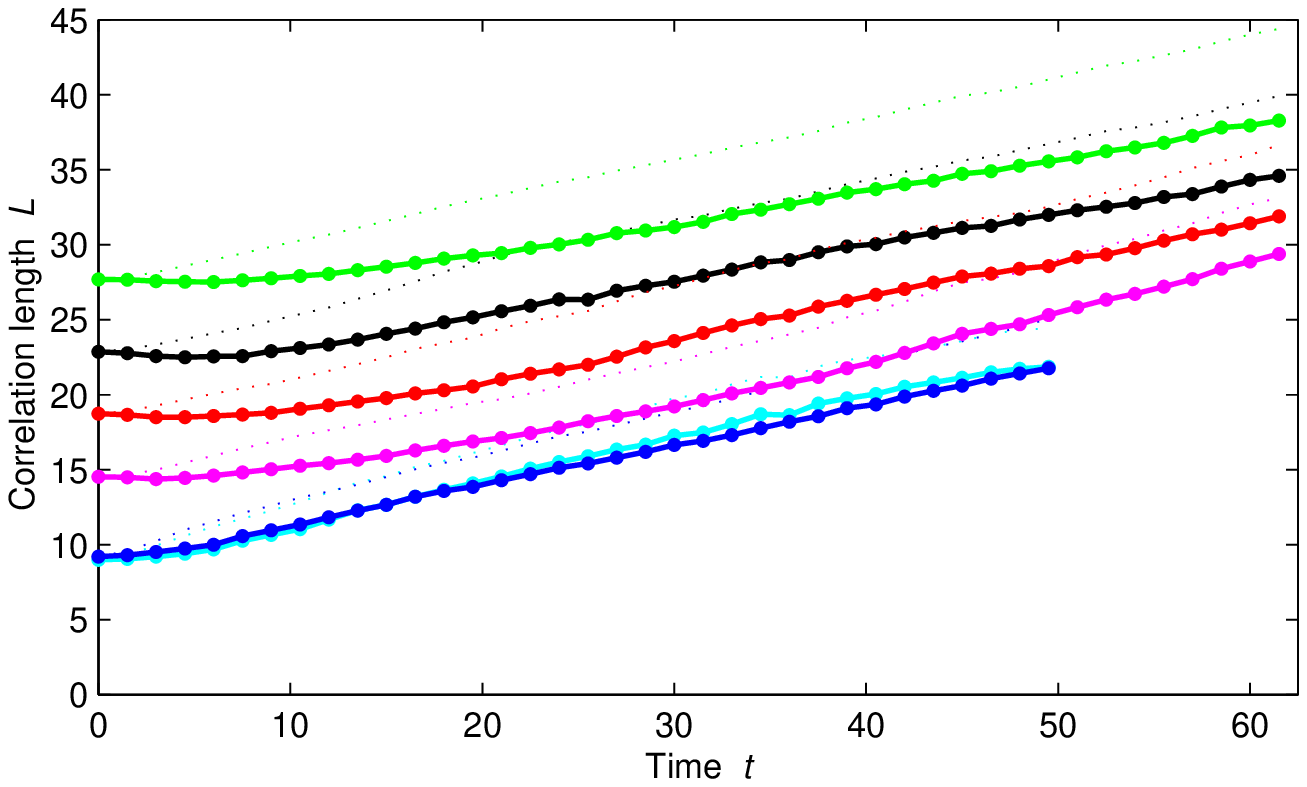}}
\centerline{
\epsfxsize=0.7\hsize\epsfbox{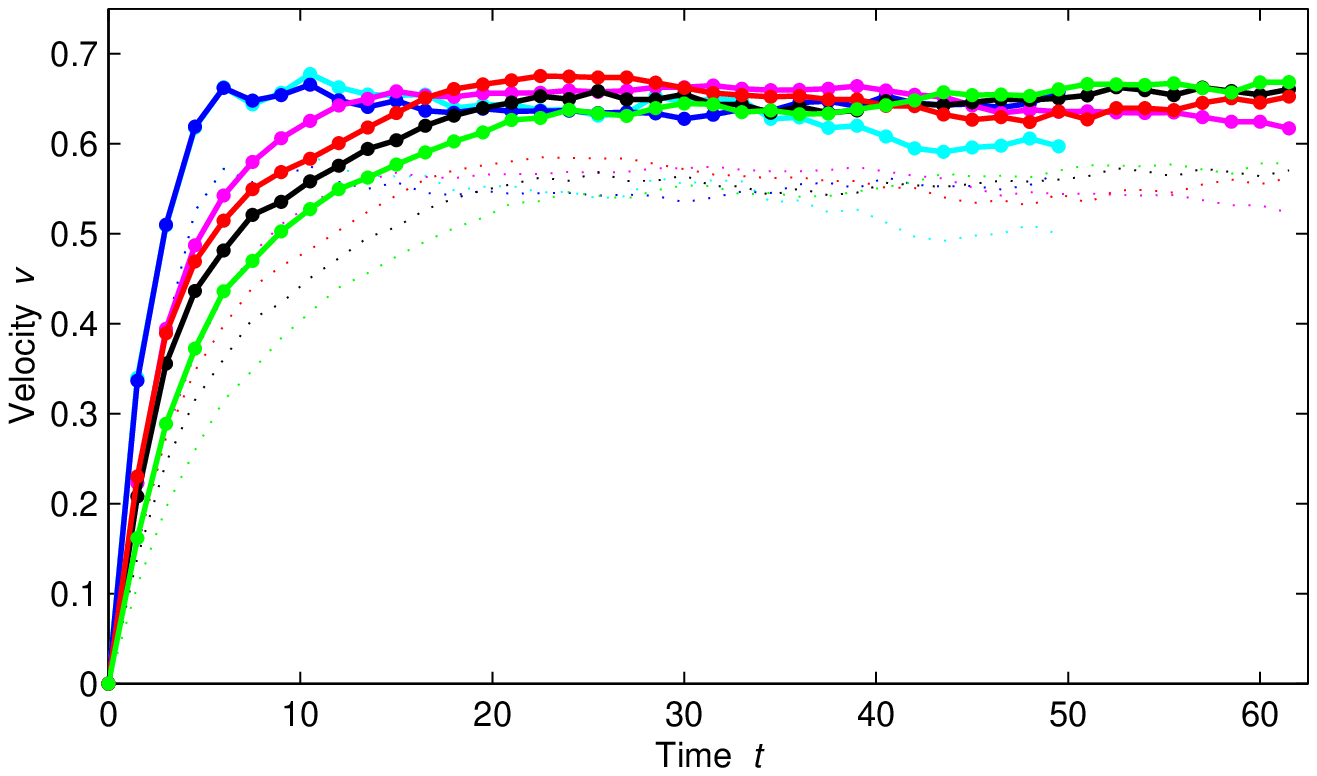}}
\vskip0in}
\caption{Flat spacetime simulation results for a series of 
simulations with a grid size of $250^3$ or above. The top panel shows 
the spatial correlation length $L$ as a function of
time. The dotted lines give the static correlation length $L_{\rm s}$ 
inferred from the static string density, rather than the total 
invariant string length used for $L$. The lower panel shows
the corresponding average rms velocity $v$ as a function of
time (with 
the same colour coding).
The dotted lines give the raw velocity before applying the velocity 
corrections plotted in Fig.~\protect\ref{figv_correction}.}
\label{figflat_data}
\end{figure}

The growth of the correlation length $L$ for flat space simulations,
along with the corresponding 
rms velocity, is illustrated in Fig.~\ref{figflat_data}.  All simulations,
which included grid sizes up to $400^3$,
were halted when the horizon size reached half the box-size. 
The solid lines in Fig.~\ref{figflat_data} represent the true
correlation length obtained
from the total string energy $L= (E/\mu)^{-1/2}$, whereas the dotted 
lines are the static correlation length $L_{\rm s}$ from \ref{static_energy}. 
This latter diagnostic, employed in \cite{Vincent1998cx},
gives much more linear initial growth and is
consistent with their results.  However, it is clear from the
velocity-dependent $L$, that the early evolution can be
interpreted as primarily due
to the acceleration of the initially static network rather than
the immediate onset of a decay mechanism capable of reducing
the invariant string length.

For higher initial densities than those plotted in Fig.~\ref{figflat_data}
there were some deviations from asymptotic linearity probably due
to the effect of background friction on string motion (this may be
evident in the bottom two curves with the highest string densities). 
The asymptotic slopes of the correlation
lengths yield the approximate scaling law
\begin{equation}
L= 0.23(\pm 0.05)t\,.
\end{equation}
Remarkably, this is close to
the result $L=0.27(\pm0.05)$ quoted for the
Smith--Vilenkin simulations of Nambu networks
in \cite{Vincent1997rb}.  
Our result is consistent to the directly comparable field theory 
simulations of \cite{Vincent1998cx}, if the results from 
their static analysis $L_{\rm s} = (0.27$--$0.34)t$ are divided
by the square root of the average Lorentz factor, $\bar\gamma \approx 1.4$.
 The initial brief burst of growth in the
correlation length seen for several
runs in \cite{Vincent1998cx} is not observed 
in our simulations and may arise from the
different procedures used to set up the initial network
configurations.  (The deviations from linearity seen in their largest run 
(see Fig.~1 in  \cite{Vincent1998cx}) appears to be due to the use of 
a larger lattice spacing for which numerical friction becomes significant.)

The approach to scaling for the velocities can be observed in the lower panel of
Fig.~\ref{figflat_data}.  The timescale for this process was rapid
and roughly inversely proportional to the initial
correlation length.  The  
asymptotic values for the different simulations were 
remarkably consistent, that is, approximately
\begin{equation}
v=0.67(\pm0.05)\,.
\end {equation}
This velocity is consistent, within our conservative error estimates, 
with naive expectations for a flat space network, that is, $v = 1/\sqrt{2}$.
As expected, the relaxation time of the velocity towards its asymptotic 
value was 
roughly proportional to the initial correlation length (compare upper and
lower panels).  The very rapid over-relaxation for the smallest correlation 
length was probably due to additional accelerations from inter-string 
forces (which have an exponential fall-off at large distances).  

\subsection{Radiation era string networks}

\begin{figure}
\vbox{\centerline{
\epsfxsize=0.7\hsize\epsfbox{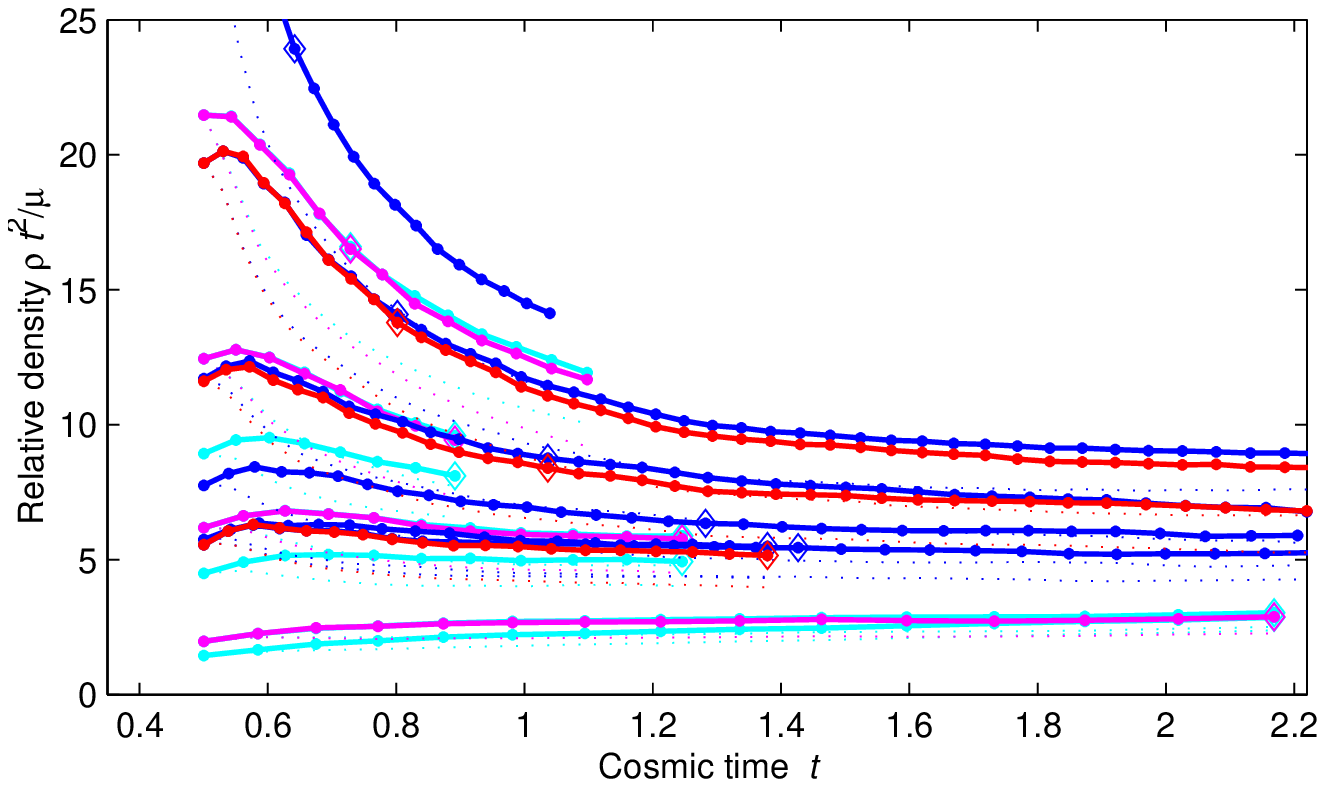}}
\centerline{
\epsfxsize=0.7\hsize\epsfbox{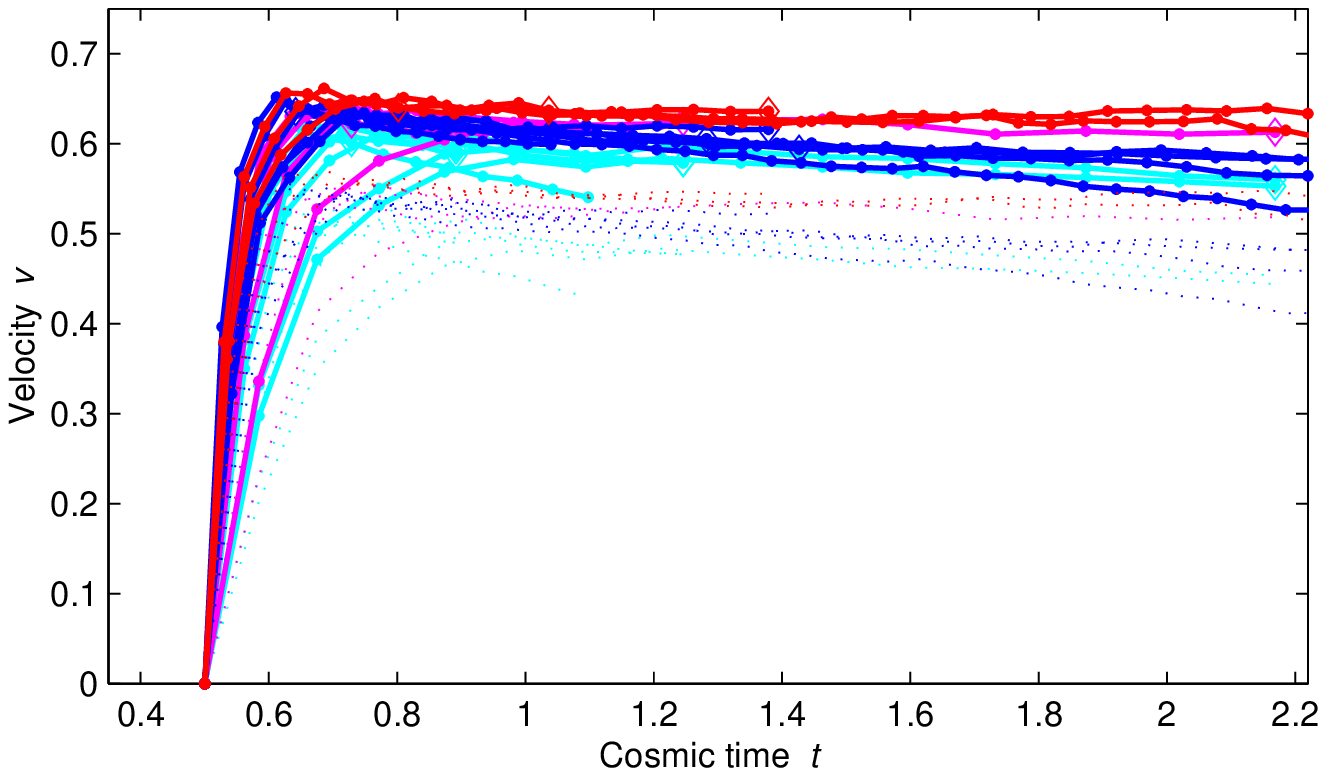}}
\vskip0in}
\caption{Radiation era simulation results:  The upper panel shows
the relative density for string networks in the
radiation era converging towards a `scaling' asymptotic value. 
The simulations
plotted in blue are large (over $320^3$) high resolution
simulations using the true physical equations of motion. These
are bracketed by smaller simulations plotted in
cyan. The simulations plotted in red are also large, but evolve
the strings using the `fat string' algorithm and these are bracketed by
small runs plotted in magenta. The dotted lines shows the string
density from a static analysis, that is, 
neglecting the Lorentz factor due to string velocities.
The lower panel shows the average rms velocity $v$ for radiation era
strings as a function of cosmic time for the series of
simulations described above (using the same colour coding). The
dotted lines show the raw measured velocity before it is
normalised.  Note the long term decline of velocities for defects 
of fixed physical size (blue).
}
\label{figrad_data}
\end{figure}

We have performed a very extensive series of network simulations 
in an expanding universe during the radiation-dominated era
$a\propto t^{1/2}$.  Simulation results for the relative energy density 
$\zeta \equiv \rho t^2/\mu $ and the rms velocity $v$ are given in 
Fig.~\ref{figrad_data}.
Small bracketing simulations are shown in cyan, with larger 
simulations ($320^3$ or above) in blue.  A diamond indicates
where the horizon size grows to half the box size.  Because 
of the consistency of the subsequent evolution we continue to 
plot it, but we note that any  analysis of these results must be 
performed
before this causal threshold is crossed.  In addition, we have
performed a series of runs using the `fat string' algorithms 
discussed earlier. The purpose of this 
was to check the relative accuracy of the two methods, rather
than to exploit the extra dynamic range that the `fat string'
method offers.  In this case, small bracketing simulations
are plotted in magenta, while the larger simulations
(again grids of $320^3$ or above) are shown in red.  There
is surprising consistency between the two methods with the 
biggest differences emerging in the late-time evolution of the 
velocities which, in turn, influences the energy density. As we
shall discuss, this is the regime in which the resolution of 
simulations with 
a physical string width becomes progressively poorer.

The asymptotic values of the relative energy density and the velocity 
were approximately
\begin{equation}
\zeta \equiv \rho t^2/ \mu = 6 (\pm 2)\,, \qquad v= 0.63 (\pm 0.05)\,.
\end{equation}
The density is considerably lower than that found in high resolution 
Nambu simulations for
which $\zeta = 13 (\pm 2)$ \cite{Bennett1990yp,Allen1990tv},
but the velocity $v=0.66$ is consistent within the uncertainties.  There is,
however, a remarkable qualitative concordance in the approach
to scaling as well as the initial relaxation.  The reason for
the lower density, such as the restricted dynamic 
range and absence of small-scale structure 
in these field theory simulations, will be discussed when we
consider the analytic modelling of these results in the next section.

\begin{figure}
\vbox{\centerline{
\epsfxsize=0.7\hsize\epsfbox{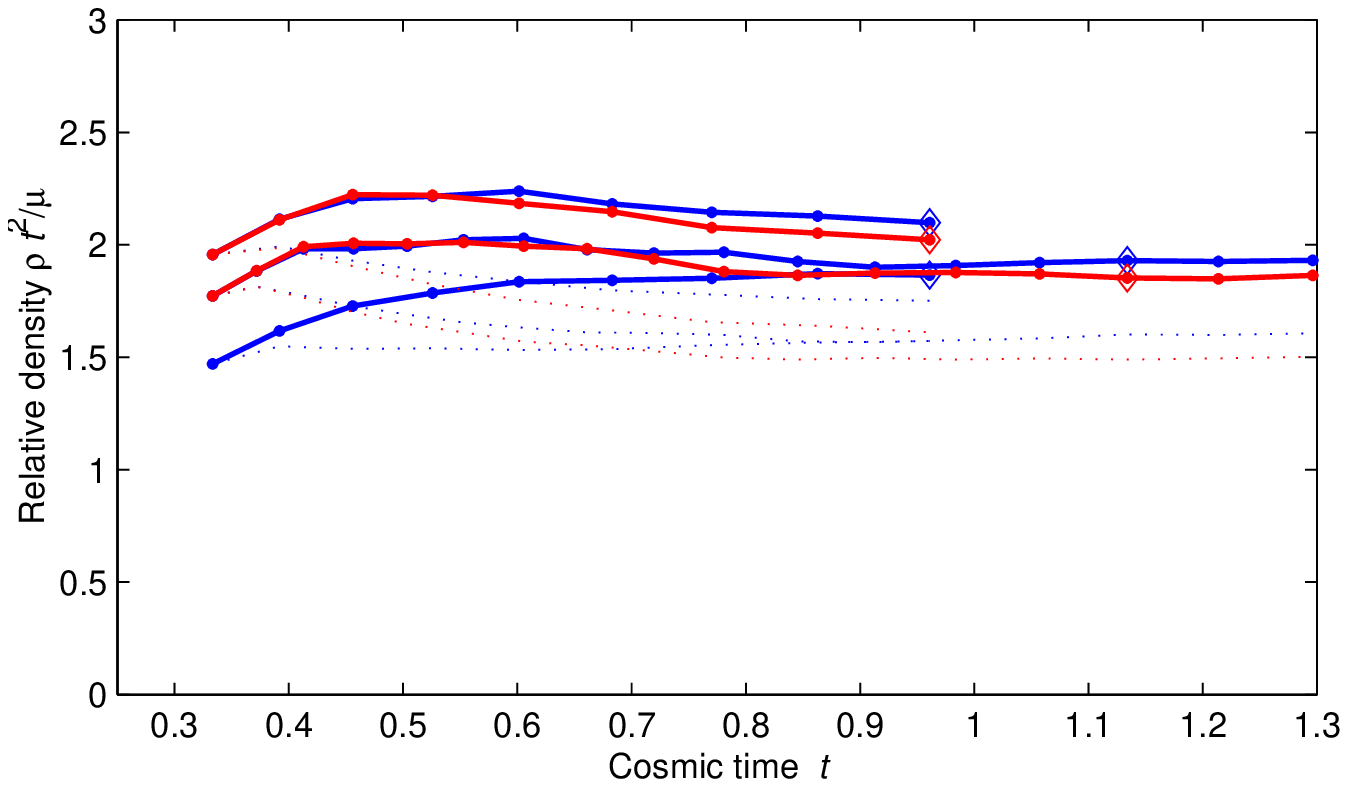}}
\centerline{
\epsfxsize=0.7\hsize\epsfbox{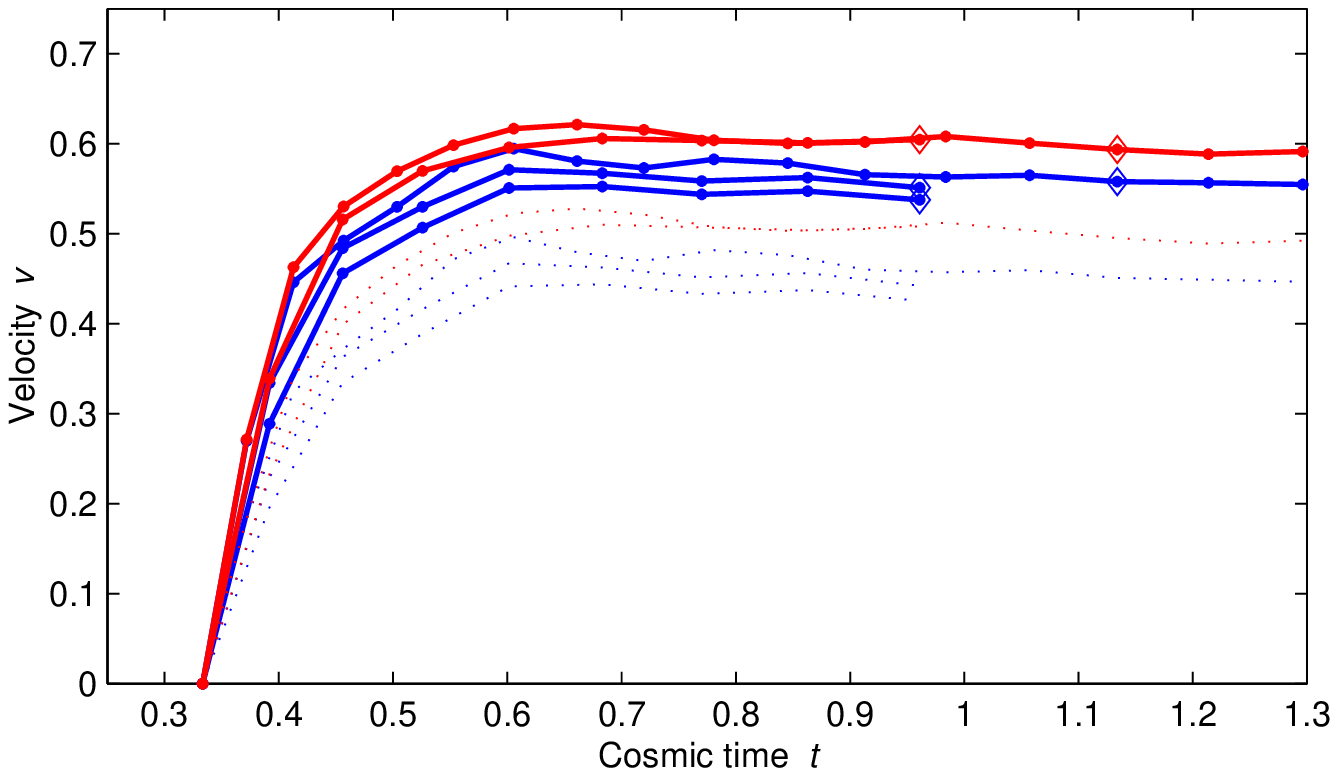}}
\vskip0in}
\caption{Matter era simulation results: The upper panel shows 
the relative density for string networks in the
matter era. The large simulations plotted in blue use the true
physical equations of motion, while those plotted in red use the
`fat string' algorithm. The lower panel shows the rms string velocity $v$ as a function of
cosmic time in the matter era (using the same colour
coding as above).}
\label{figmat_data}
\end{figure}

\subsection{Matter era string networks}

Matter era simulations shown in Figs.~\ref{figmat_data} 
were not studied in great detail, except 
to broadly bracket the apparent scaling value. Combining the simulations
we suggest approximate asymptotic values for the relative density and 
the rms velocity of
\begin{equation}
\zeta = 2.0 (\pm 0.5)\,,\qquad v= 0.57 (\pm 0.05)\,.
\end{equation}
Again, this density is roughly half that found in high resolution 
simulations for Nambu networks where $\zeta = 3.5(\pm 1)$, but the 
velocity is again consistent with 
$v = 0.58$--$0.61$ \cite{Bennett1990yp,Allen1990tv}.  Regardless of
the overall velocity normalisation, the relative velocities for 
flat space, radiation and matter eras were found to be in the 
ratios $1.0:0.93:0.83$ which is remarkably consistent with Nambu
simulation ratios $1.0:0.94:0.85$.  Clearly, Hubble damping is acting
on the extended field theory vortex-strings in a similar manner to 
the one-dimensional Nambu strings.

With the rapid expansion occurring in the matter era, the defect
width shrinks very quickly on a comoving grid 
and so the effective dynamic range of field theory simulations
is severely restricted.  After reaching the threshold where the defect
width and lattice spacing were comparable, numerical accuracy declined
dramatically and the velocity plummeted towards zero.  In this respect,
the `fat string' algorithm fared very much better (plotted in red)
 showing no sign of the initial decline in 
velocities evident in simulations solving the true physical equations
of motion.

\subsection{Global string networks}

\begin{figure}
\vbox{\centerline{
\epsfxsize=0.7\hsize\epsfbox{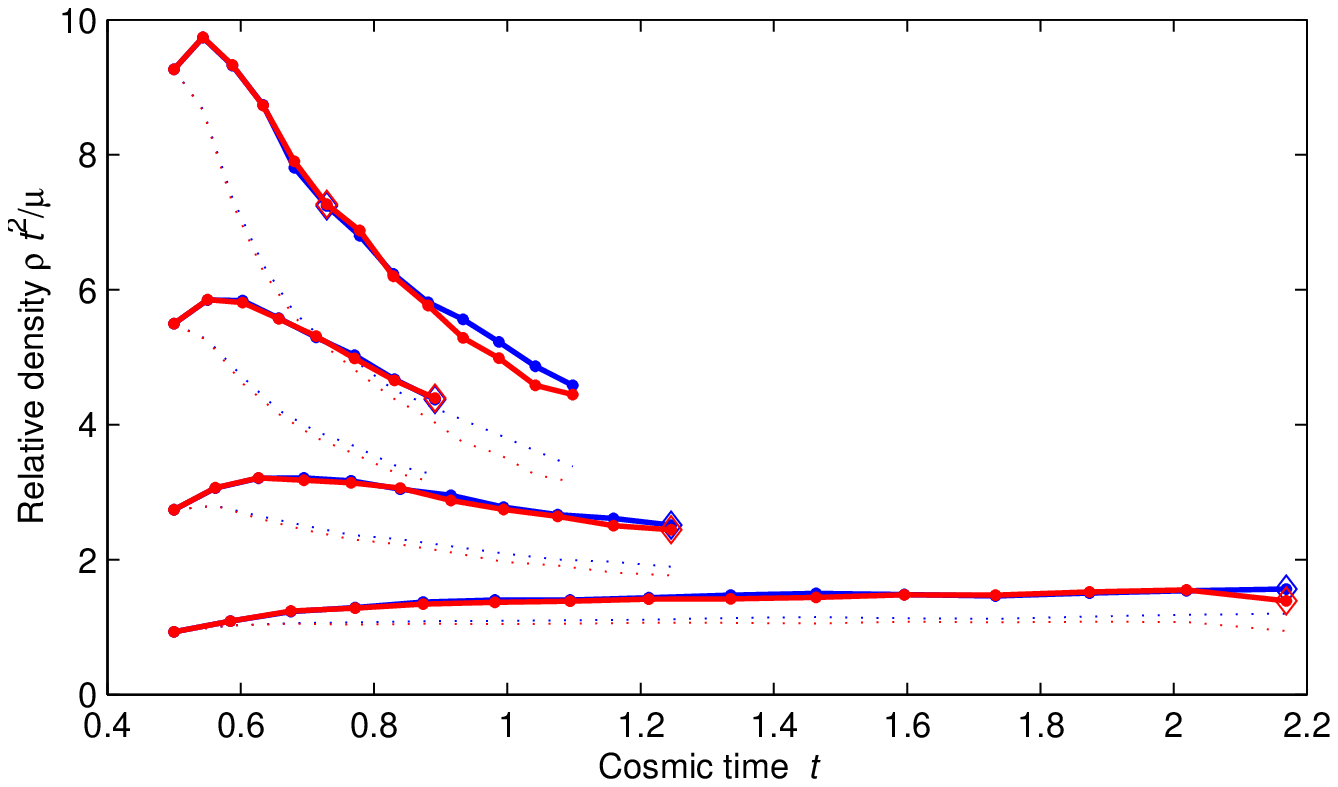}}
\centerline{
\epsfxsize=0.7\hsize\epsfbox{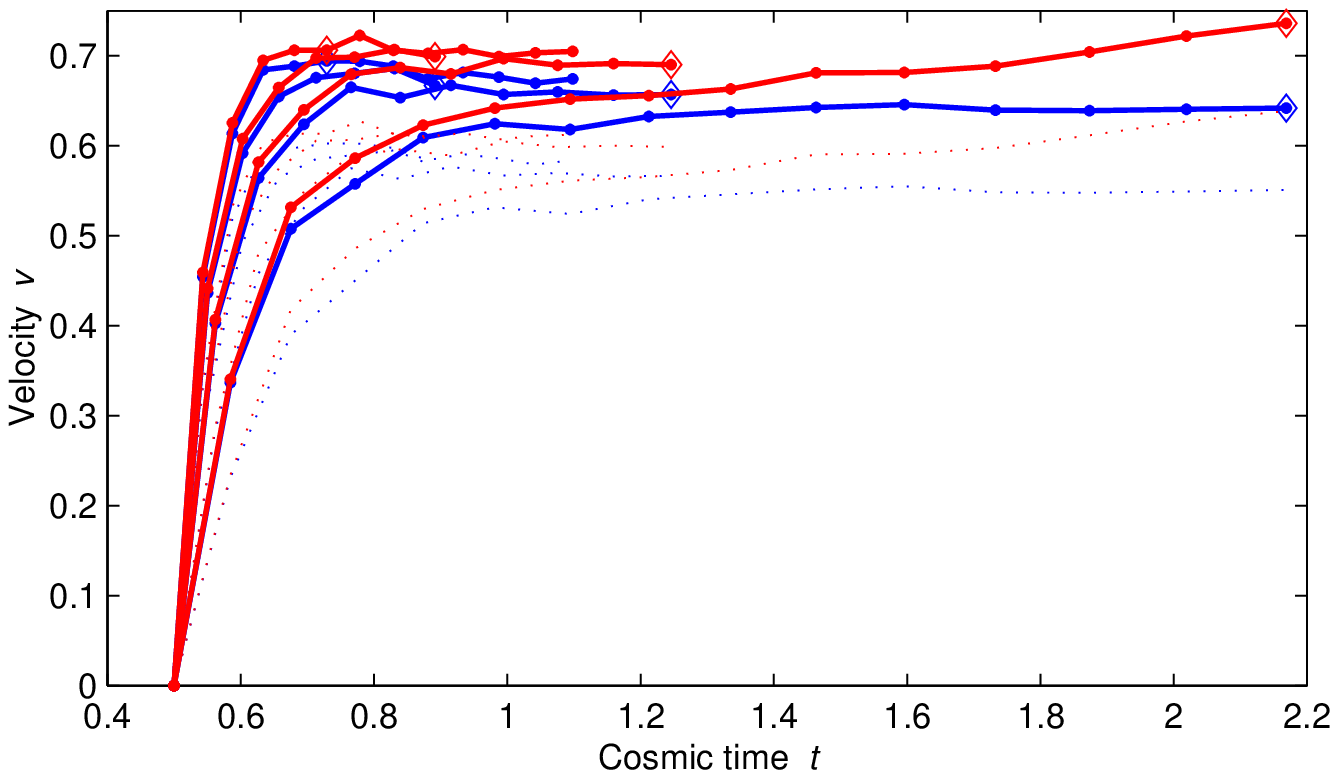}}
\vskip0in}
\caption{Global string simulations in the radiation era:
The upper panel shows relative density for global string networks in
the radiation era. The large simulations plotted in blue use the
true physical equations of motion, while those plotted in red
use the `fat string' algorithm.
The lower panel shows the average rms velocity $v$ as a function of
time for global strings in the radiation era (using the same
colour coding as above). Note the diverging velocities for the 
true (blue) and `fat' (red) global strings.
}
\label{figglobal_rad_data}
\end{figure}

Global strings provide an interesting contrast to gauged strings because
they are known to be strongly radiating, especially on the small scales
probed by these simulations.  As a further independent check on our work, 
 in \cite{Yamaguchi1999yp,Yamaguchi1999dy} global string networks  
have also been simulated in both matter and 
radiation eras.  Here, we study global 
strings using the same \AH{} code but with 
the couplings altered appropriately.  The relative scaling density and
velocity were found to be, respectively (refer to 
Figs.~\ref{figglobal_rad_data}), 
\begin{equation}
\zeta = 2.0 (\pm 0.5)\,,\qquad v= 0.68 (\pm 0.05)\,.
\end{equation}
(Note that we use the defects of fixed physical width (blue) to estimate
the velocity.)  
We believe this result for $\zeta$ is consistent with the lower result $\zeta = 0.9$--$1.3$ 
quoted in ref.\cite{Yamaguchi1999yp} because
the two Lorentz factor corrections to 
their static analysis (discussed previously), would make the total 
invariant string
density higher by a factor of two.

With only about 8 global strings crossing each horizon volume, this is 
a dramatically lower density than the 24 strings observed in the local
gauged case (and the 50 strings seen in the Nambu simulations). 
Clearly, the global strings, through their coupling to the 
massless Goldstone boson, have available a much more effective energy 
loss mechanism on these scales than local strings which only couple to 
massive fields.  Despite this fact, however, radiation back-reaction 
effects do 
not reduce the average rms string velocity, instead it appears to be 
systematically higher than the local case.   
This indicates that the strong long-range interactions 
between global strings are playing an important role in accelerating
the network dynamics.

Finally, we note that the `fat string' algorithm does not appear to mimic
the true string dynamics as effectively in the global case (compare the
red velocity curves in Fig.~\ref{figglobal_rad_data} with the radiation era
local strings, Fig.~\ref{figrad_data}).
The physical and `fat string' velocities diverge significantly 
even when the numerical evolution should remain accurate (that is,  before
horizon crossing of the box). 
An important physical effect mitigates against the accuracy of the `fat 
string' algorithm in the global case.  The string energy density 
(and the inter-vortex potential) has a logarithmic term, $\mu \propto 
\ln (L/\delta)$, with a width cutoff $\delta$ which continues to grow 
in this case.  This enlarged cutoff $\delta$ implies a lower effective mass
and so the string will
accelerate more rapidly under the long-range inter-string forces present, and 
it will also radiate its energy more efficiently.

\subsection{Efficacy of the `fat string' algorithm}

Summarising our previous discussions, we conclude that our new 
`fat string' algorithm for local strings in an expanding universe, 
appears to closely replicate the averaged properties of true string
network dynamics.  Both the string densities and velocities were in
good agreement over the limited dynamic range available to this study. 
This encourages further application of this method to field theory 
simulations of local strings in an expanding background, thereby
accessing much larger dynamic ranges.  However, the detailed implications
of the `fat string' algorithm deserve further study, especially for
the evolution of the Hamiltonian, the propagation of massive radiation, 
and the emergence of small-scale structure on strings.  As we have also
seen, the `fat string' algorithm for global strings fared satisfactorily, 
but did not reproduce string velocities as successfully.  It does not 
provide a very precise quantitative representation of global string
evolution over the small dynamic ranges probed by these simulations,
let alone those required to make accurate cosmological predictions.

\section{Analytic modelling of network evolution}
\label{secanalytic}

\subsection{The velocity-dependent one-scale model}

The velocity-dependent one-scale (VOS) model has been described in
considerable detail elsewhere
\cite{Martins1995tg,Martins1996jp,Martins1997thesis,Martins2000cs},
so here we limit ourselves to a brief discussion, highlighting the
features that will be important for what follows. This generalized
 `one-scale' model \cite{Kibble1985hp}, 
aims to describe the general evolutionary properties
of the string network through the behaviour of a small number of
averaged or `macroscopic' quantities, namely its energy $E$ and rms
velocity $v$ defined respectively by
\begin{equation}
  E=\mu a(\tau)\int\epsilon d\sigma\, , \qquad
  v^2=\frac{\int{\dot{\bf x}}^2\epsilon d\sigma}{\int\epsilon
    d\sigma}\,,
  \label{eee}
\end{equation}
where the string trajectory ${\bf x}(\sigma, t)$ is parametrised by
the world-sheet coordinates $\sigma$ and $t$ and the `energy density'
$\epsilon (\sigma,t)$ gives the string length per unit $\sigma$
along the string.  For coherence, here, we summarise the main points
of the original `one-scale' model in preparation for the new ingredients of
the VOS model.  

The long string network is a Brownian random walk on large scales
and can be characterised by a correlation length $L$. This can be
used to replace the energy $E = \rho V$ in long
strings in our averaged description, that is,
\begin{equation}
\rho_\infty \equiv \frac{\mu}{L^2}\,.
\label{corr_length}
\end{equation}
A phenomenological term must then be included to account for the
loss of energy from long strings by the production of loops, which
are much smaller than $L$. A `loop chopping efficiency' parameter
$\tilde c$ is introduced to characterise this loop production as
\begin{equation}
  \left(\frac{d\rho_{\infty}}{dt}\right)_{\rm to\ loops}=
  {\tilde c}v_\infty\frac{\rho_{\infty}}{L}
  \, . \label{rtl}
\end{equation}
In this approximation, we would expect the loop parameter $\tilde c$
to remain constant irrespective of the cosmic regime, because it is
multiplied by factors which determine the string network
self-interaction rate.

From the microscopic string equations of motion, one can then
average to derive the evolution equation for the correlation length
$L$,
\begin{equation}
  2\frac{dL}{dt}=
  2HL(1 + {v_\infty^2})+
  \frac{L}{\ell_{\rm f}}{v_\infty^2}
  +{\tilde c}v_\infty \, , \label{evl0}
\end{equation}
where $H$ is the Hubble parameter and $\ell_{\rm f}$ is a friction
damping length scale. The first term in (\ref{evl0}) is due to the
stretching of the network by the Hubble expansion which is modulated
by the redshifting of the string velocity. The second term is due to
frictional interactions by a high density of background particles
scattering off the strings. The friction length scale $\ell_{\rm f}$
(defined in \cite{Martins1995tg}) typically depends on the
background temperature as $\ell_{\rm f} \approx {\cal O}(1)\mu
T^{-3}$, so that it grows with the scale factor as $a^{3}$ and it
will become irrelevant at late times. It proves easiest (and
physically clearer) to characterise this friction term by a
parameter $\theta$ which measures the initial ratio of the damping
terms due to friction and expansion. In the case of flat spacetime,
however, this friction length scale will be a constant, and hence the
effect of this term, if at all relevant, can not be neglected at
late times.

The effect of massless radiation on the long-string network can be
included in the evolution equation for the correlation length
(\ref{evl0}) in the same way as previously achieved for the
evolution of the length of a string loop \cite{Martins1996jp}. Taking
for example the typical case of gravitational radiation, one easily
finds \cite{Martins2000cs} that the following term can be added to the
right-hand side of (\ref{evl0})
\begin{equation}
  2\left(\frac{dL}{dt}\right)_{\rm massless}\equiv8\Sigma v^6_\infty\, ; \label{grb}
\end{equation}
Here, $\Sigma$ is a constant which is the long-string counterpart of $\Gamma 
G\mu$ found for the gravitational decay of string loops ($\Gamma \approx 65$)
\cite{Vilenkin1994cossot}.
We note that a term for Goldstone boson radiation should also take this form,
but with a much larger coefficient $\Sigma$ than for gravitational
radiation \cite{Battye1994qa}.
For the radiation of massive particles (mass $m$) we expect 
the above term to be exponentially suppressed (or similar) 
for correlation lengths well
beyond an appropriate inverse mass-scale $L_{\rm d}\sim m^{-1}$.  Hence, we can 
introduce an analogous phenomenological term to describe it:
\begin{equation}
  2\left(\frac{dL}{dt}\right)_{\rm massive}\equiv8\Sigma v^6_\infty\exp
  \left(-\frac{L}{L_{\rm d}}\right)\, . \label{grb2}
\end{equation}

One can also derive an evolution equation for the long string
velocity with only a little more than Newton's second law
\begin{equation}
  \frac{dv_\infty}{dt}=\left(1-{v^2_\infty}\right)
  \left[\frac{k}{L}-\left(2H+\frac{1}{\ell_{\rm f}}\right)v_\infty\right]\, ,
  \label{evv0}
\end{equation}
where $k$ is called the `momentum parameter'. The first term is the
acceleration due to the curvature of the strings and the second
damping term is from both the expansion and background friction. The
parameter $k$ is defined by
\begin{equation}
  k\equiv\frac{\langle(1-{\dot {\bf x}^2})({\dot {\bf x}}\cdot {\bf
      u})\rangle}
  {v(1-v^2)}\, ,
  \label{klod}
\end{equation}
with ${\dot {\bf x}}$ the microscopic string velocity and ${\bf u}$
a unit vector parallel to the curvature radius vector.  An accurate
ansatz for $k$ can be derived and justified---see  \cite{Martins2000cs}. 
For strings in near vacuum, we are interested in the relativistic 
regime  for which a sufficiently precise phenomenological 
form is 
\begin{equation}
  k_{\rm rel}(v) =\frac{2\sqrt{2}}{\pi}\;\frac{1-8v^6}{1+8v^6}\, .
  \label{krel2}
\end{equation}
We note that in the opposite friction-dominated case, the non-relativistic
limit is $ k_{\rm nr}(v)={2\sqrt{2}}/{\pi}$.
The non-relativistic scaling
law predicted is then
$L\propto t^{1/2}$ and $v\propto t^{-1/2}$ \cite{Martins1996jp}.

The VOS model model has been extensively
compared with the results of numerical
simulations \cite{Bennett1990yp,Allen1990tv,Martins2000tesvdo} and
shown to provide a good fit to the large-scale properties of a
string network. In particular, it matches well the evolution between
asymptotic regimes as a network passes through the matter--radiation
transition. Comparisons with numerical simulations confirm the
constancy of the only free parameter, the loop chopping efficiency
$\tilde c$, and fix its value in the expanding universe to be \cite{Martins2000tesvdo}
\begin{equation}
  \tilde c = 0.23 \pm 0.04\,,
  \label{cpar}
\end{equation}
and, in Minkowski space, 
\begin{equation}
  \tilde c = 0.57 \pm 0.04\,,
  \label{cparflat}
\end{equation}
that is, about twice as large (we will return to this point below).  
For a detailed
discussion of other regimes of applicability of this model (such as
open and $\Lambda$-universes), as well as an analysis of its various
scaling solutions, refer to \cite{Martins2000cs}.

In what follows, we will use this model to fit the numerical
simulations described previously. We will concentrate on trying to
identify which of the different dynamical mechanisms (loop
production, friction or radiation) is dominant for the evolution of
the network.  In doing so we will look for distinguishing features 
that differentiate between these mechanisms.  For example, friction 
reduces string velocities, whereas loop production and radiative
effects exhibit different behaviours in their initial relaxation 
due to their different dependence on the velocity $v$.  Note that
we will only endeavour to fit data before the horizon crosses the
numerical box, that is, for times prior to the diamonds plotted in
the expanding universe simulations. 

\subsection {Flat space modelling}

\begin{figure}
\vbox{\centerline{
\epsfxsize=0.7\hsize\epsfbox{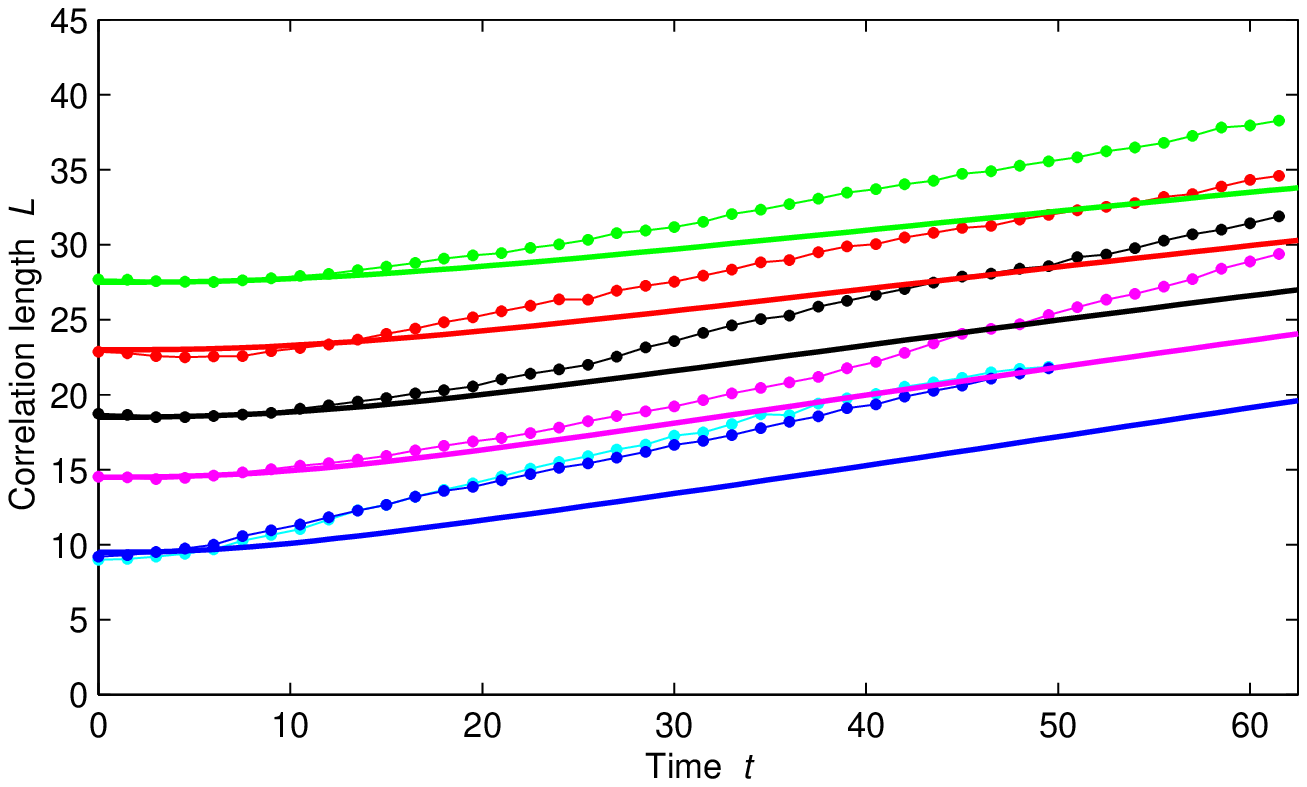}}
\centerline{
\epsfxsize=0.7\hsize\epsfbox{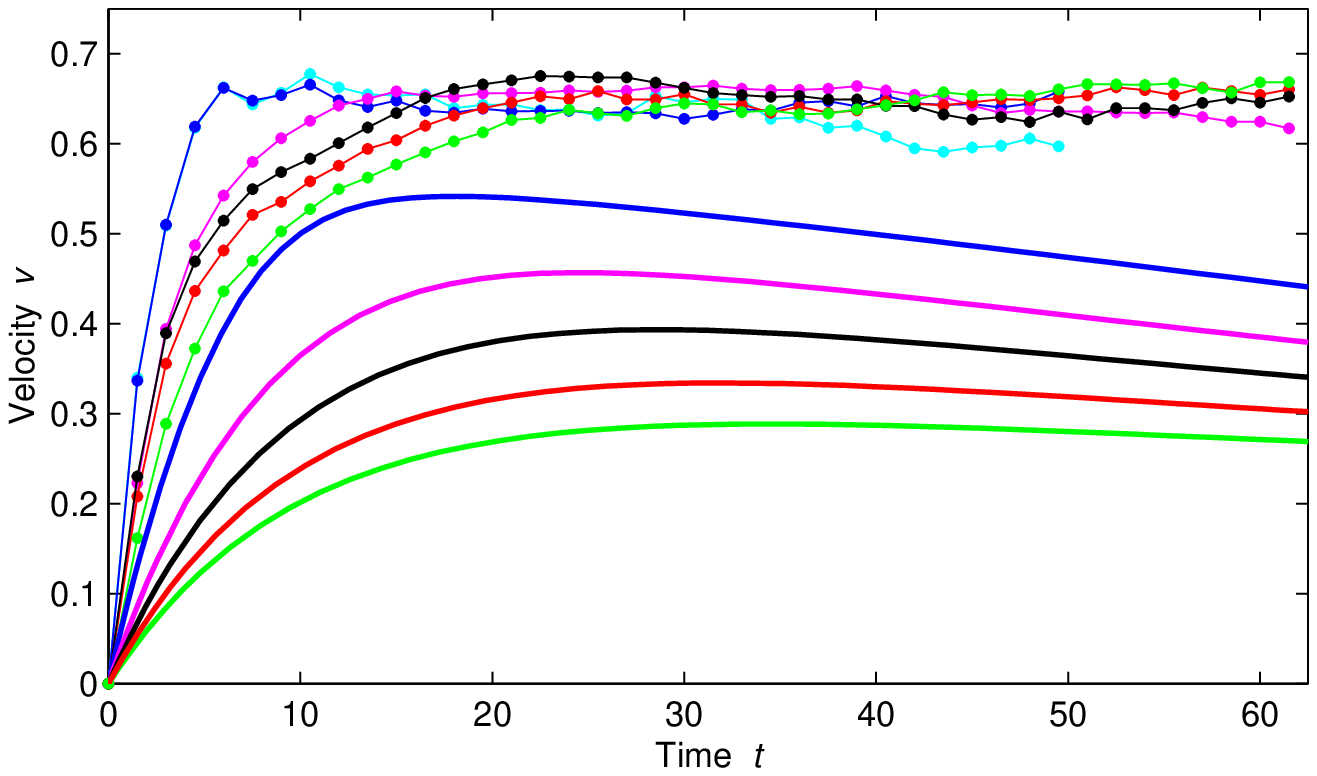}}
\vskip0in}
\caption{Friction-dominated strings in flat space.  The
best fit VOS model with friction only 
$\theta=0.1$ (solid lines), is compared to the 
flat space simulation data (data points).  These fits are very poor,
especially for the velocities,  and rule out friction as a dominant
dynamical effect in these string network simulations.}
\label{flat_vos_fric}
\end{figure}

We start by considering the effect of friction alone, that is,  
we hypothesise that the initial gradient flow actually leaves 
behind a substantial background between the relaxed strings in
the initial configuration. Friction has two competing effects 
on the string density (or correlation
length). On one hand, increasing the friction parameter $\theta$
will make $L$ grow faster, but on the other hand it will also
decrease the string velocity $v$. Given that the friction term is
proportional to $Lv^2$, there will be an `optimal' value of $\theta$
that will provide a best-fit for the string density. We find that
this is near $\theta\sim0.1$, and we compare with simulations in 
Fig.~\ref{flat_vos_fric}. We can see that it is a very poor fit, primarily 
because the velocities are too low and their asymptotic values
differentiate (unlike the consistent asymptote of the 
simulation data). For this reason, 
we can rule out friction as an important influence on the network 
dynamics, giving us greater confidence in the effectiveness of 
the initial gradient flow in creating a negligible background.

\begin{figure}
\vbox{\centerline{
\epsfxsize=0.7\hsize\epsfbox{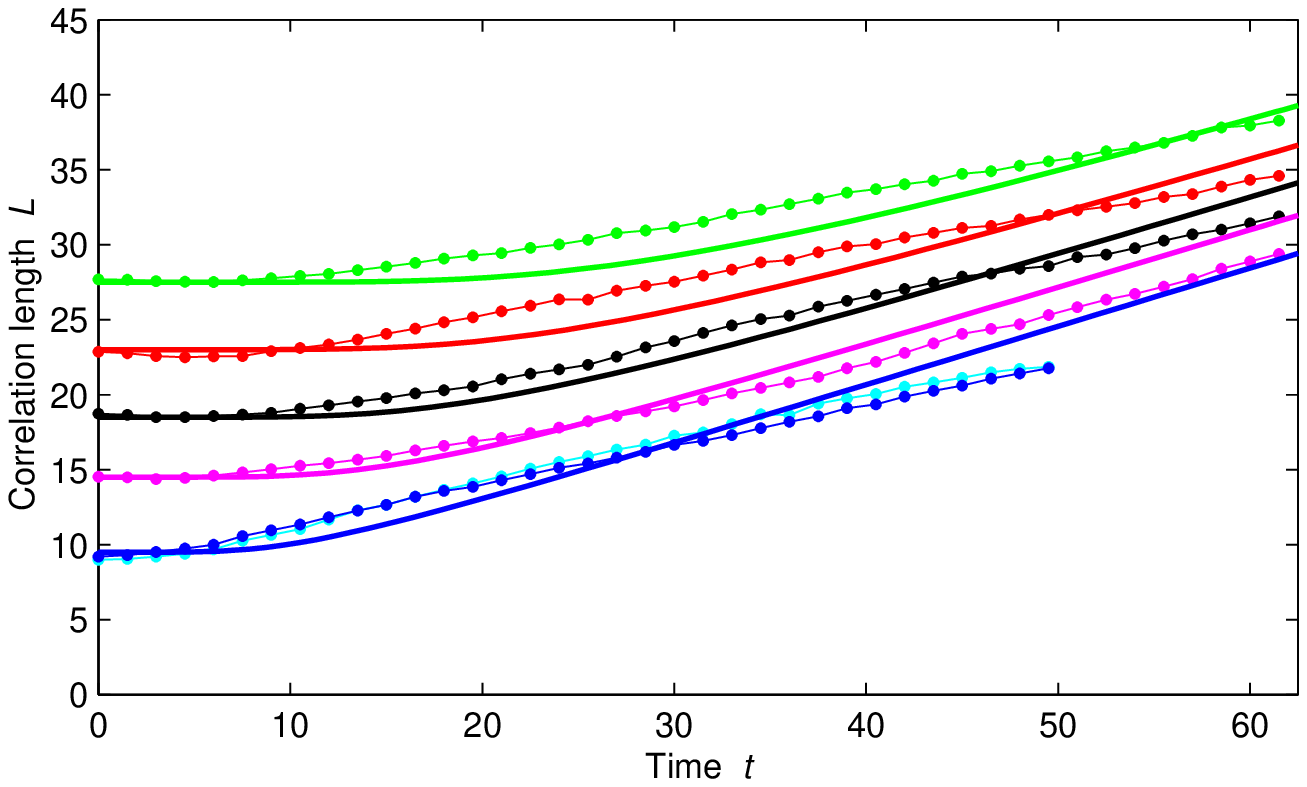}}
\centerline{
\epsfxsize=0.7\hsize\epsfbox{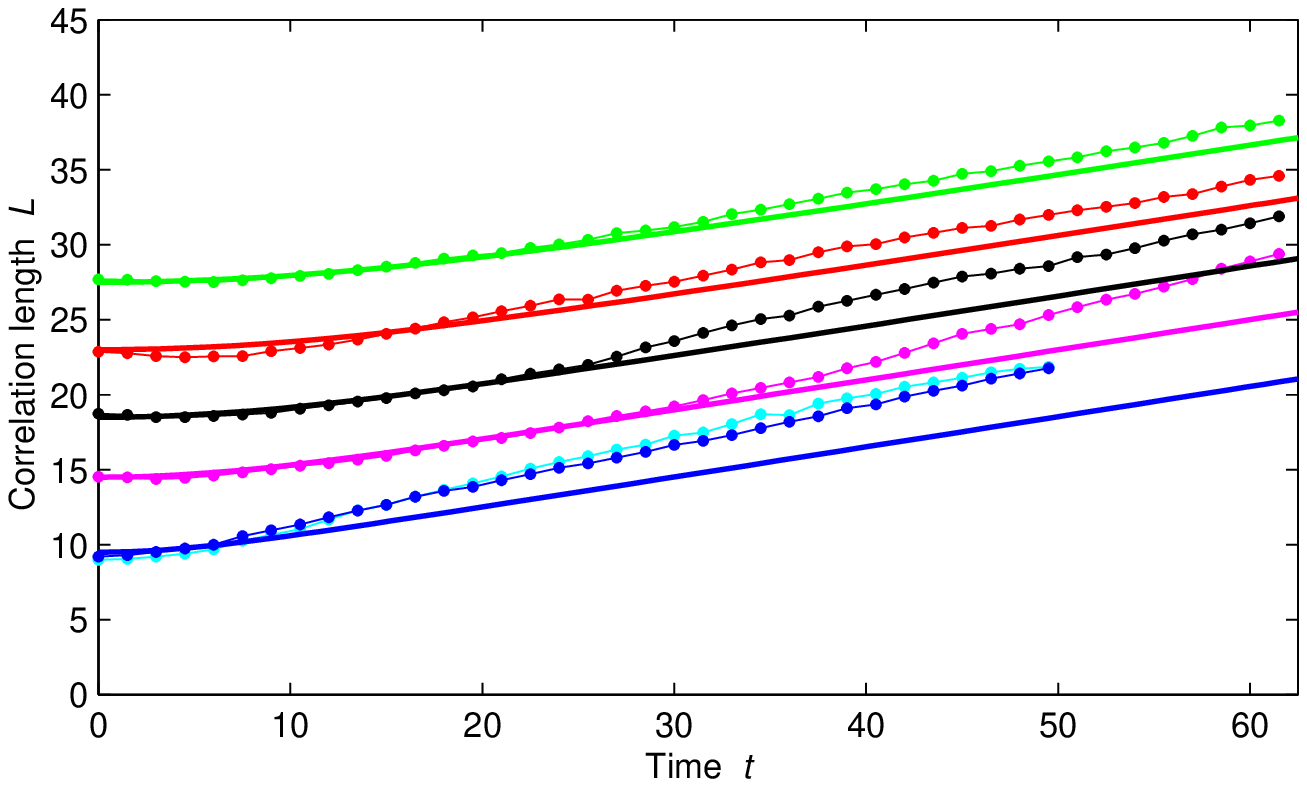}}
\vskip0in}
\caption{Possible energy loss mechanisms for flat space 
string networks showing analytic fits
for a `scale-invariant massive radiation' scenario (upper panel)
and for the standard loop production model (lower panel). 
In the upper panel, the VOS model (solid lines) with only `massless' 
radiation $\Sigma=0.8,
L_{\rm d}\rightarrow \infty$, ($\tilde c=0$)  is a satisfactory 
fit to the simulation data points.
In the lower panel, the best fit (solid lines)
for loop production has ${\tilde c}=0.57$ ($\Sigma=0$) which
corresponds to the value also found for flat space Nambu networks. 
There is some evidence of small deviations from the simulated 
data (solid lines) for small initial correlation lengths $L$.
}
\label{flat_vos_damp_loop}
\end{figure}

We next consider the effect of radiation in the VOS model,
with a best-fit shown in the upper panel of 
Fig.~\ref{flat_vos_damp_loop} for the radiation
parameter $\Sigma=0.8$ ($L_{\rm d}\rightarrow \infty$).
This is motivated by the `scale-invariant massive radiation' scenario 
proposed in  ref.~\cite{Vincent1998cx} and in which massive particle emission
is {\it not} cut-off beyond some inverse mass-scale ($L_{\rm d}\rightarrow \infty$).
Although our previous work suggests that massive radiation cannot be produced 
in this manner \cite{Moore1998gp}, there is a reasonable fit in 
Fig.~\ref{flat_vos_damp_loop} to the flat
space simulation data, implying we cannot rule out this possibility
on these grounds alone. However, we note the particular parameter choice
with $\Sigma=0.8$ and will confront the model later with expanding universe data.

Next we consider the case of loop production only, see the lower panel of 
Fig.~\ref{flat_vos_damp_loop}.
Here the best fit is provided by ${\tilde c}=0.57\pm0.05$. Note that
this is precisely the same value that was found in flat spacetime
Nambu string simulations \cite{Martins2000cs,Martins2000tesvdo}
(refer also to \cite{Vincent1997rb}).
We can see that the fit is quite good for initial conditions
corresponding to high values of the correlation length, and that it
gets comparatively less accurate as this decreases. This is
what one would expect if there is an extra, relatively
small initial contribution from massive radiation.

\begin{figure}
\vbox{\centerline{
\epsfxsize=0.7\hsize\epsfbox{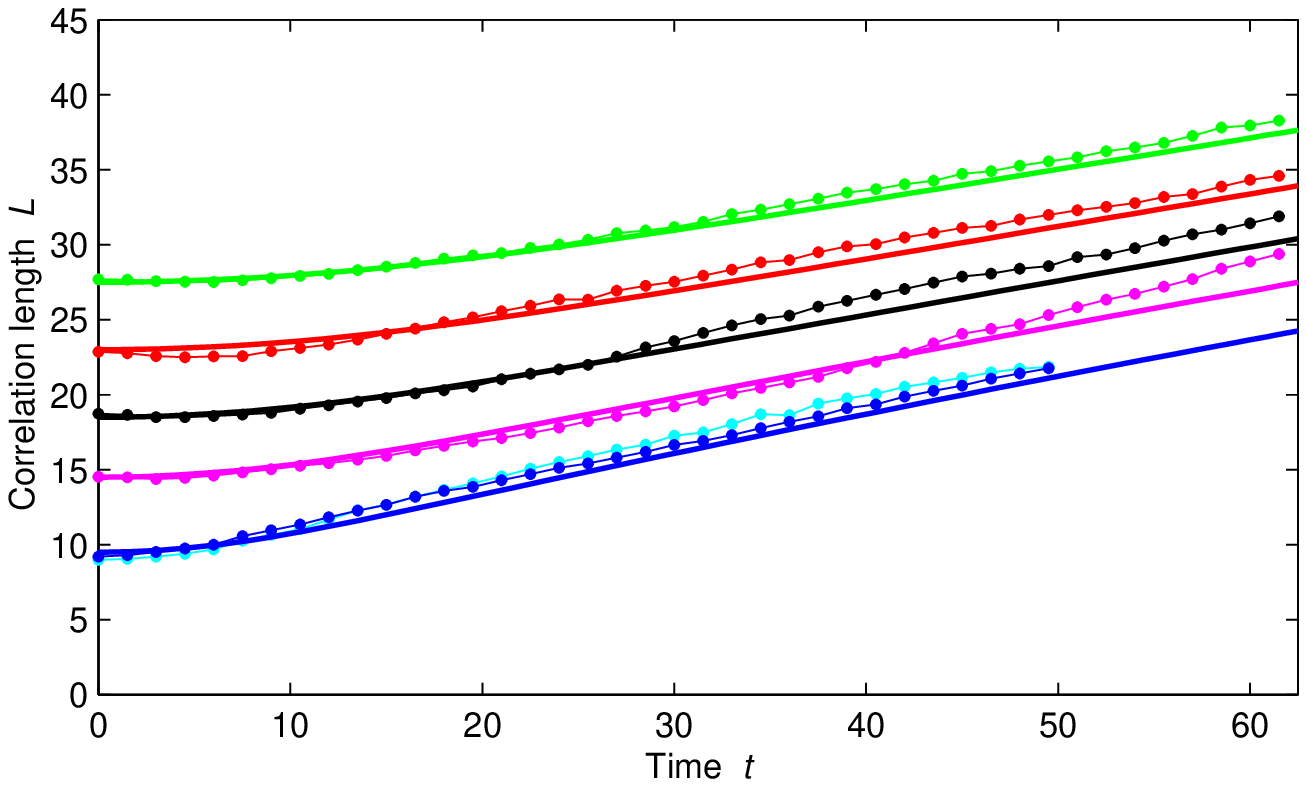}}
\vskip.4in}
\vbox{\centerline{
\epsfxsize=0.7\hsize\epsfbox{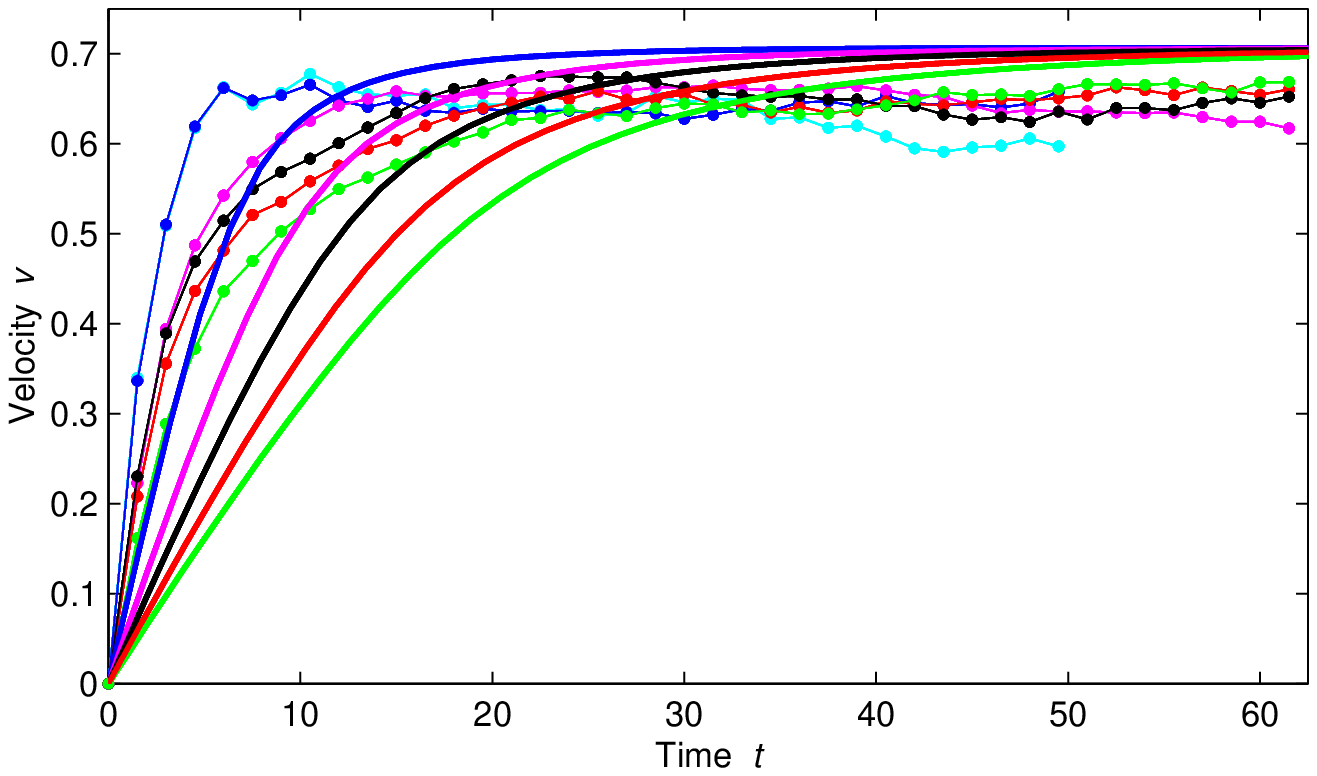}}
\vskip.4in}
\caption{Loop production model with some massive radiation 
for flat space string networks.  This shows the very good 
fit between simulated
strings (solid lines) and the VOS model (dashed lines) assuming
loop production (${\tilde c}=0.57$) with some massive radiation 
($\Sigma=0.5, ~L_{\rm d}=4\pi$). }
\label{flat_vos_best}
\end{figure}

In Fig.~\ref{flat_vos_best} we show an improved fit where we have added
a massive radiation term of strength $\Sigma =0.5$ and the exponential 
cut-off value $ L_{\rm d} = 4\pi$ (motivated by our study of massive
radiation from strings \cite{Moore2000a}).  This choice implies that 
loop production is predominant throughout this simulation but is 
supplemented with a brief initial burst of massive radiation if the 
correlation length is sufficiently small (refer to (\ref{grb2})). 
One can readily observe that  that this provides a very good 
fit to the correlation lengths and to the asymptotic velocities. 
(Indeed, this is our overall `best fit' model in that it is the parameter choice which 
provides good agreement for all thirty different local string simulations, 
whether 
in flat space or an expanding universe.)  Here, we appear to have 
found  a rather encouraging correspondence between the
present flat spacetime field theory simulations, analogous
Nambu simulations and the analytic VOS model. 

\begin{figure}
\vbox{\centerline{
\epsfxsize=0.7\hsize\epsfbox{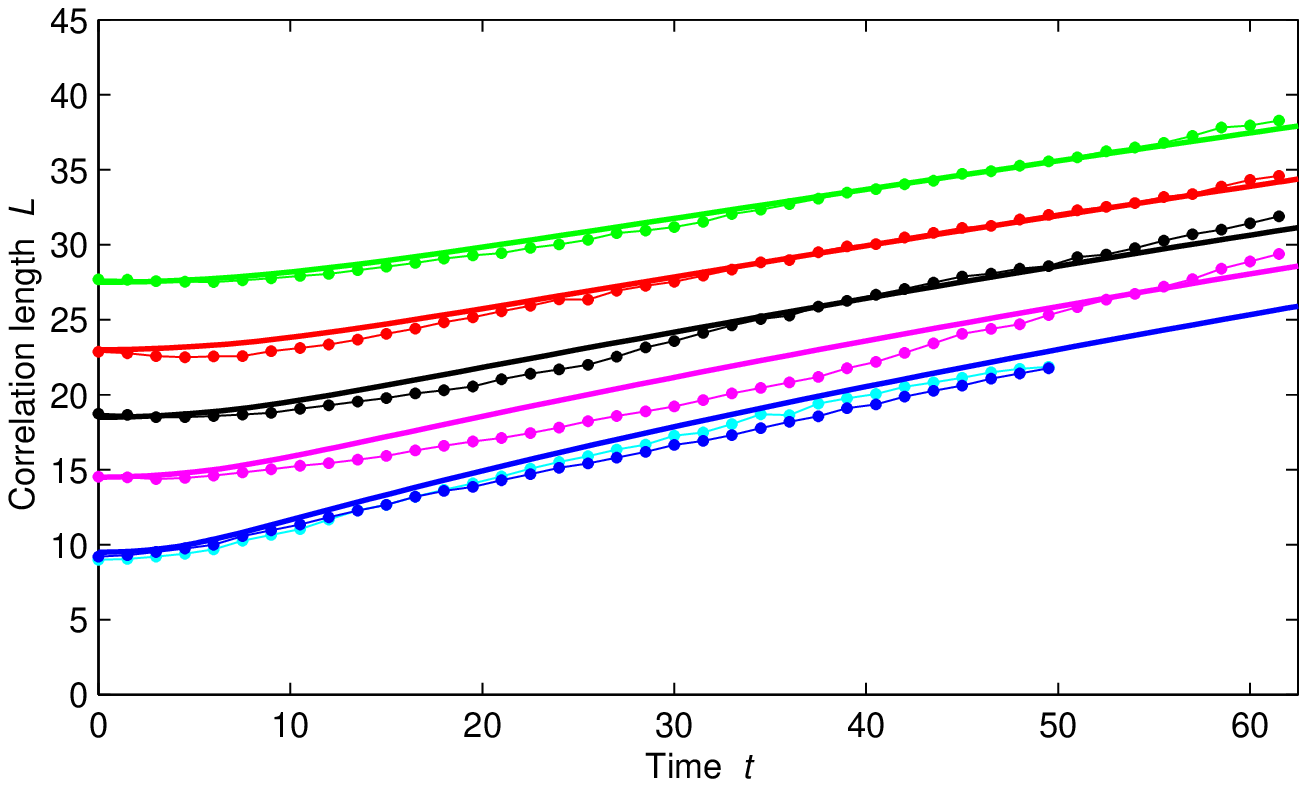}}
\vskip.4in}
\vbox{\centerline{
\epsfxsize=0.7\hsize\epsfbox{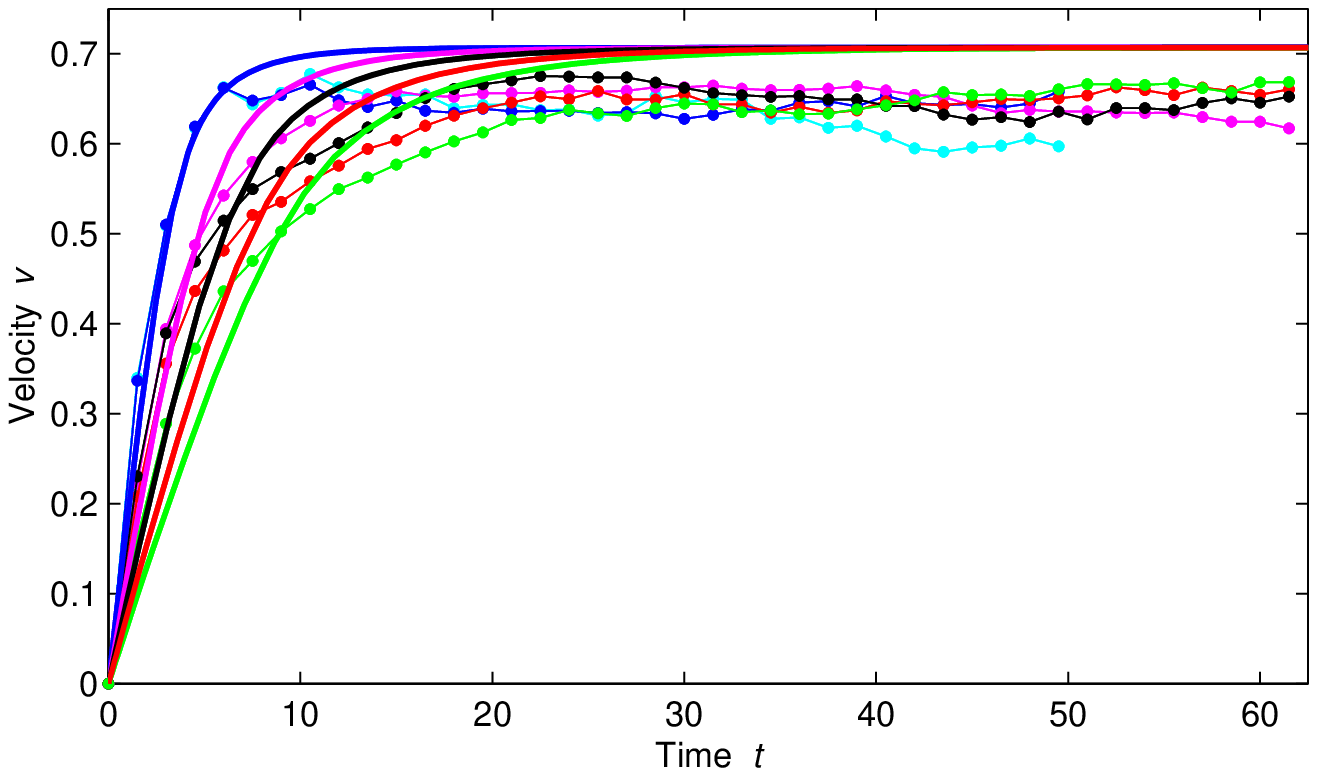}}
\vskip.4in}
\caption{An alternative fit to the flat space simulation data
using a `corrected' velocity VOS model with the radius of 
curvature $R= L/2$ (see text).  Here, the parameter choice is
${\tilde c}=0.45$, $\Sigma=1$ and $L_{\rm d}=4\pi$.}
\label{flat_vos_alt}
\end{figure}

\subsection{An alternative velocity fit}

The one shortcoming of the correspondence between the simulation data
and the VOS model is in the fast initial rise of the velocities, before
they begin to approach their asymptotic values. This is not entirely
surprising, because the initial conditions can be chosen such that 
the string curvature radius $R$ and the string correlation
length $L$ (as defined in (\ref{corr_length})) 
are significantly different;  the VOS model
assumes that they are comparable. In principle, a phenomenological term 
could incorporate this effect
by modifying the $k/L$ term in the VOS model (\ref{evv0}).
We find that an initial ratio
of $L\sim 2 R$ would significantly improve the early-time velocity
fits. The best-fit for the flat spacetime simulations is
shown for comparison in  Fig.~\ref{flat_vos_alt};
note that the best-fit model parameters are now ${\tilde c}=0.45$ 
and $\Sigma=1$. Even though this improves the very early time fits
(notably for the velocities), it does not 
do as well in terms of key asymptotics.  Indeed, to obtain a 
satisfactory fit in the expanding case we require a rather different
parameter choice (for radiation $\tilde c = 0.4$, $\Sigma = 0.75$).

Interestingly, this factor of two correction between the curvature radius 
$R$ and correlation length $L$ is confirmed when they are measured
directly from the simulation initial conditions. However, as 
the simulation evolves,
the initial conditions are erased and the two length scales
become increasingly similar. Thus there is nothing
`fundamental' about this initial ratio of 2---it is related
to the particular way in which the network is generated using 
diffusive evolution. Other initial conditions, such as those from 
the Vachaspati-Vilenkin
algorithm,
produce statistically different networks which would have
different $L/R$ ratios (in this case 1.4), which 
evolve towards unity in a different way.  We, therefore, disregard 
the transient effect of the initial conditions and continue 
with the unmodified VOS model because of its success in matching
important asymptotic results in both flat and expanding backgrounds.

We note also that the VOS model is missing any small-scale physics 
beyond the accelerations due to the string curvature.   Inter-string forces
are significant for the small string separations inherent in our particular
initial conditions.  The strongest evidence for their influence has 
already been discussed for global strings.

\subsection{Expanding universe modelling}

\begin{figure}
\vbox{\centerline{
\epsfxsize=0.7\hsize\epsfbox{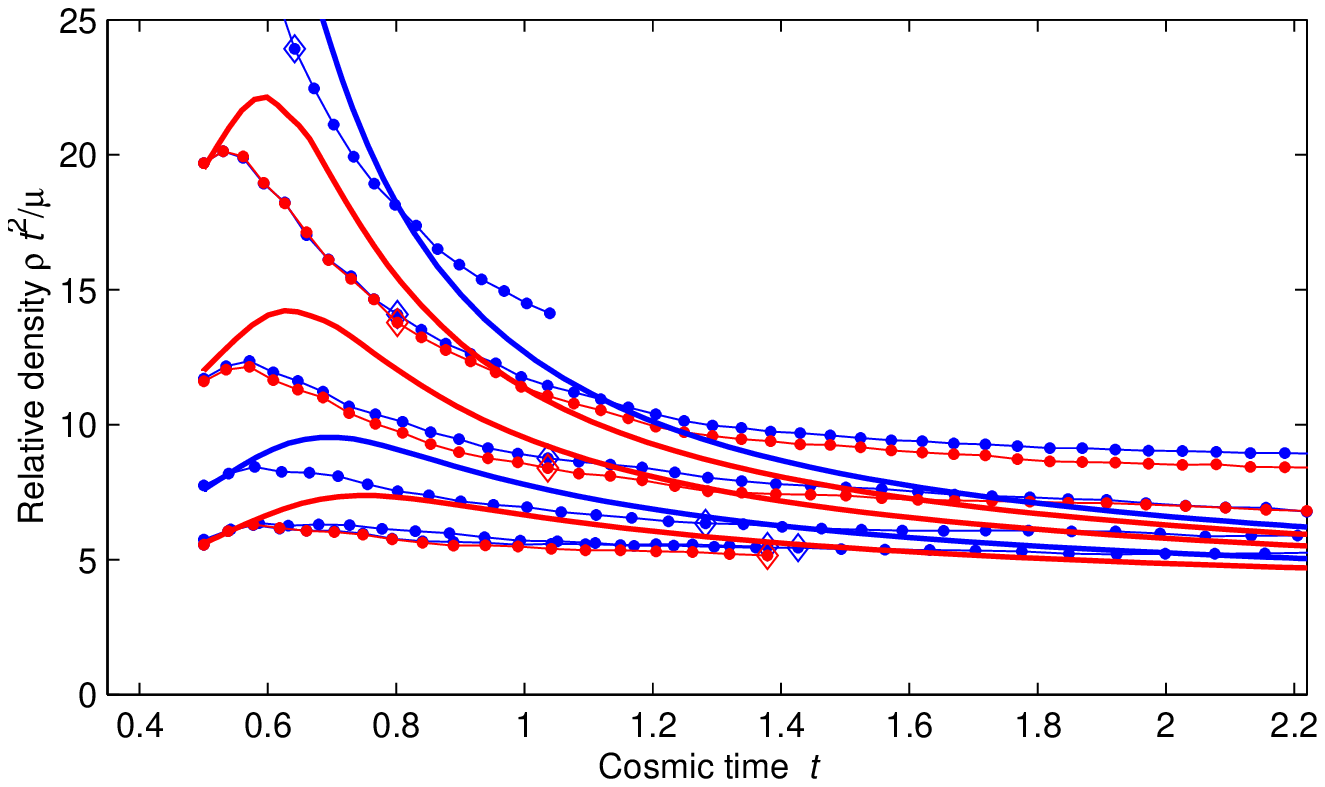}}
\centerline{
\epsfxsize=0.7\hsize\epsfbox{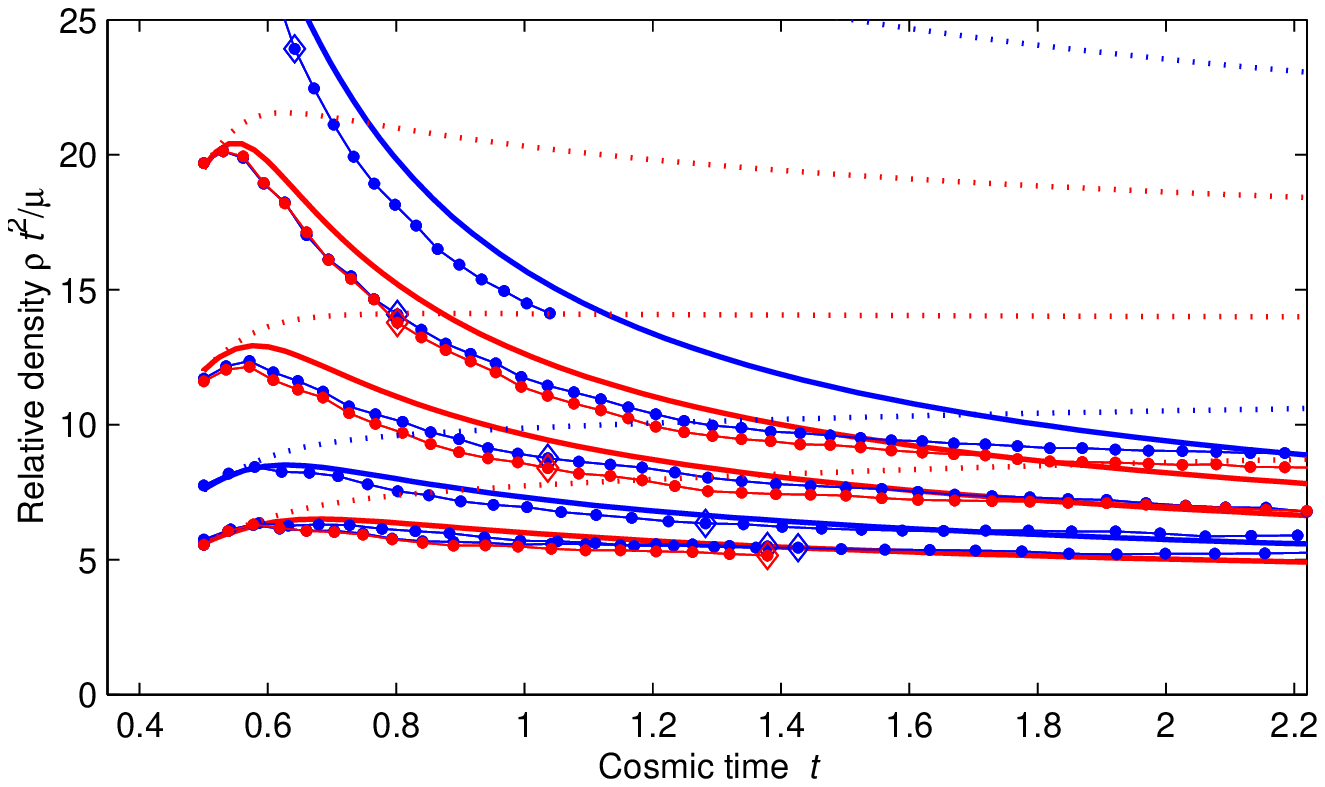}}
\vskip0in}
\caption{Top panel: `Scale-invariant massive radiation' scenario
for radiation era strings.  The best fit VOS model (solid lines)
with `scale-invariant massive radiation' only, 
$\Sigma=0.8, L_{\rm d}\rightarrow \infty$ ($\tilde c = 0$) 
produces a
poor initial fit to the simulation data (red and blue lines).
Bottom panel: Loop production model in the radiation era. 
The overall best fit VOS model (solid lines) 
has the same loop production coefficient
${\tilde c}=0.57$ ($\Sigma =0$)  as flat space simulations.
Using ${\tilde c}=0.23$ found for radiation era Nambu networks 
produces a poor fit (dotted lines) to the simulated data.}
\label{rad_vos_damp_loop}
\end{figure}

We begin for the radiation era data by considering the individual
effects of massive radiation and loop production, 
refer to Fig.~\ref{rad_vos_damp_loop}.
The `scale-invariant massive radiation' scenario is shown in the top panel;
the best-fit value $\Sigma=0.8$ (assuming $L_{\rm d}\rightarrow \infty$), 
is the same as in the flat spacetime case.
However the fit to the energy density appears to be rather poor, primarily
because the initial relaxation is too slow due to the inherent
velocity dependence of the radiation terms in (\ref{grb2}). It appears that the
`scale-invariant massive radiation' scenario of ref.~\cite{Vincent1998cx}
does not provide an adequate description of the actual network dynamics.

In the bottom panel of Fig.~\ref{rad_vos_damp_loop} we have plotted the 
model predictions
for both ${\tilde c}=0.57$ (appropriate in flat spacetime 
for both \AH{} and Nambu string networks) and for
${\tilde c}=0.23$, which was found
to provide the best fit for
both radiation and matter era Nambu 
simulations\cite{Martins2000cs,Martins2000tesvdo}.
We find that  ${\tilde c}=0.57$ provides a very reasonable fit 
(solid lines), whereas
${\tilde c}=0.23$ clearly does not (dotted lines). This is, however, not
entirely unexpected given that Nambu network simulations can probe a
much wider range of length scales below the correlation length, thus allowing
small-scale `wiggles' to build up on those scales. In other words,
no currently available field theory simulation has a spatial resolution
or dynamic range sufficiently large to allow for the build-up of
small-scale structures on the strings.

The value of
${\tilde c}=0.57$ can therefore be regarded as a `bare' loop
chopping efficiency, while ${\tilde c}=0.23$ can be interpreted as
a `renormalised' one. Note that this is also consistent with the fact
that Nambu simulations in
the expanding case somewhat surprisingly possess much more 
small-scale structure than corresponding flat
spacetime strings (for example, as quantified by the fractal properties
of each network) \cite{Martins2000cs,Martins2000tesvdo}. 
Note also that the approximate factor of two difference
between the two loop production rates may be related to the
well-known result that
the `renormalised' and `bare' string mass per unit length differ by
about a factor of two in radiation era Nambu 
simulations \cite{Bennett1990yp,Allen1990tv,Martins2000tesvdo}.
We will discuss these issues further in the next section.    

\begin{figure}
\vbox{\centerline{
\epsfxsize=0.7\hsize\epsfbox{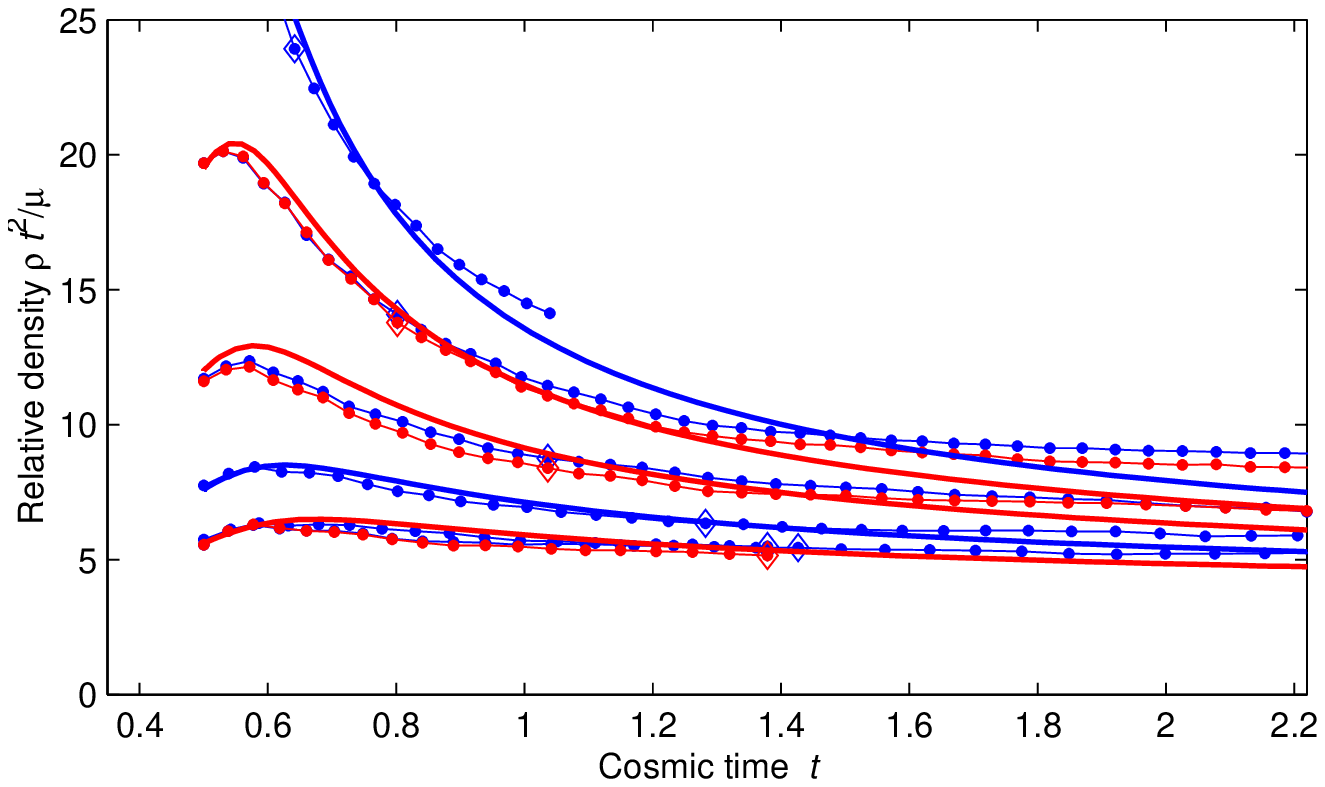}}
\centerline{
\epsfxsize=0.7\hsize\epsfbox{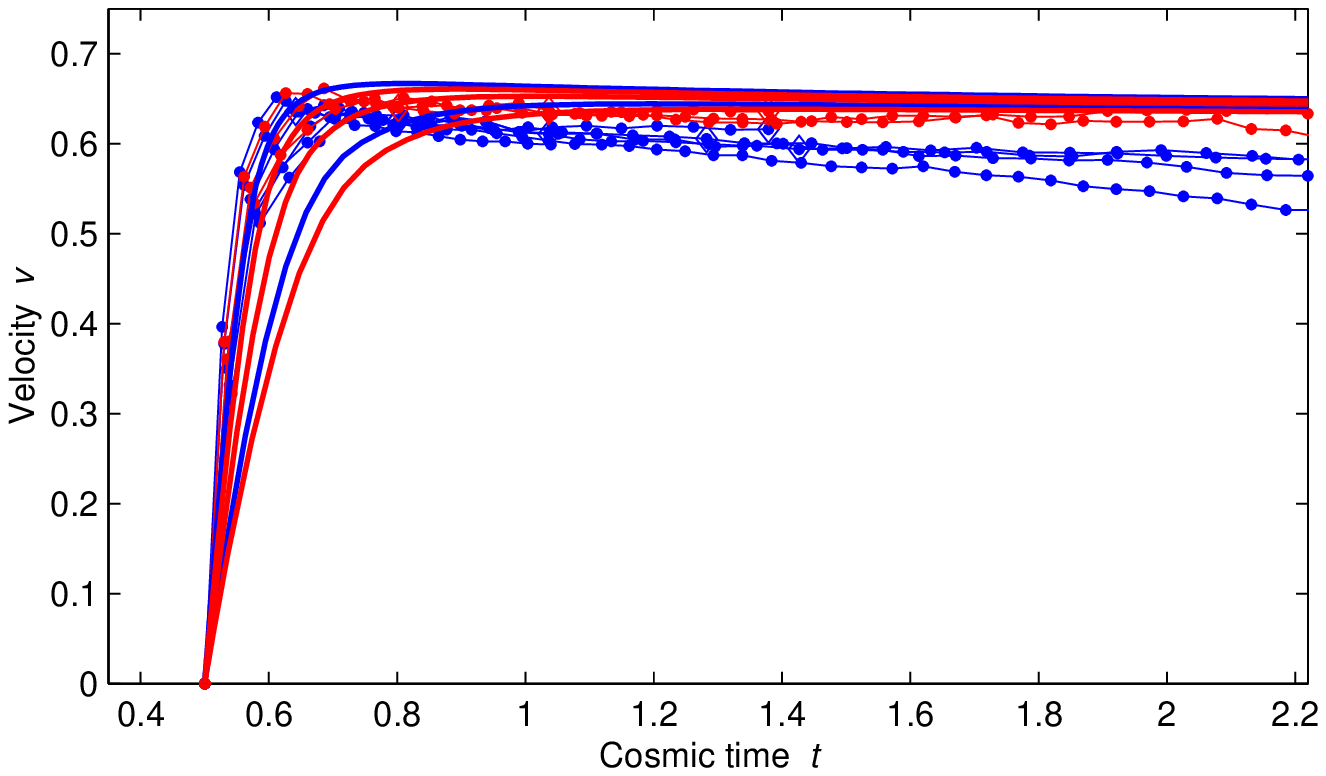}}
\vskip0in}
\caption{Radiation era simulation results fitted with our standard 
`best fit' VOS model, that is, loop production model with some massive
radiation.  This excellent fit 
has the same parameters as for the flat space results in 
Fig.~(\ref{flat_vos_best}), that is, 
${\tilde c}=0.57$, $\Sigma=0.5$, and $L_{\rm d}=4\pi$ (appropriately rescaled). 
Note, as in other figures, that comparisons must be made before the horizon 
crosses the numerical box (marked with a diamond).}
\label{rad_vos_best}
\end{figure}

In Fig.~\ref{rad_vos_best}, 
we provide what appears to be an excellent 
fit to the radiation era data using exactly the same model parameters as
in flat space.  The model is 
loop production supplemented with some massive radiation again using  
${\tilde c}=0.57$, $\Sigma=0.5$. This surprisingly good fit implies that
we have found remarkable consistency between the flat spacetime
and radiation era simulations using what can be described as the 
standard picture of string network evolution. 

\begin{figure}
\vbox{\centerline{
\epsfxsize=0.7\hsize\epsfbox{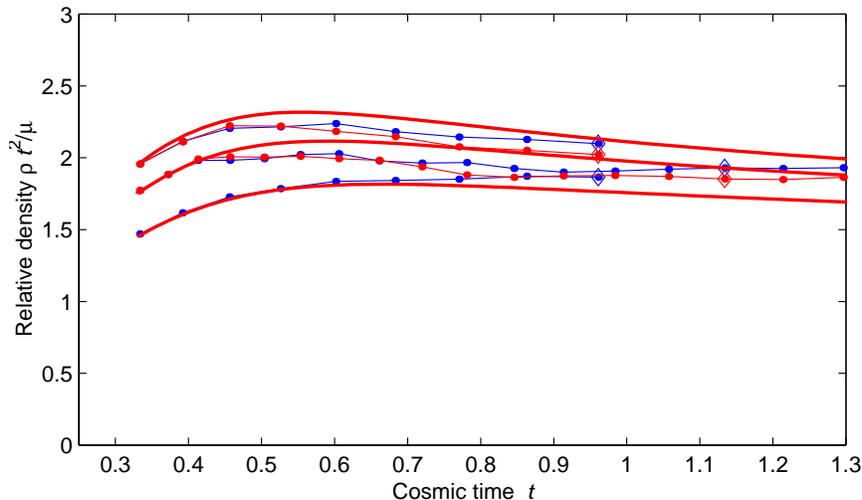}}
\vskip.4in}
\caption{Analytic modelling of matter era string simulations.
Here we find a good fit to the simulated data points
using the VOS model (solid lines) with loop production and some
massive radiation using radiation era/flat space parameters
(${\tilde c}=0.57, \,\Sigma=0.5$, and $L_{\rm d}=4\pi$).}
\label{mat_vos_best}
\end{figure}

For completeness we also compare  the VOS model with the limited 
simulation data we currently have available for string networks in the matter
era.  Fig.~\ref{mat_vos_best} illustrates a very satisfactory correspondence
between the data and the 
loop production/massive radiation description which does so well in 
the radiation era and flat space (again ${\tilde c}=0.57$, $\Sigma=0.5$).

\subsection{Global string modelling}

\begin{figure}
\vbox{\centerline{
\epsfxsize=0.7\hsize\epsfbox{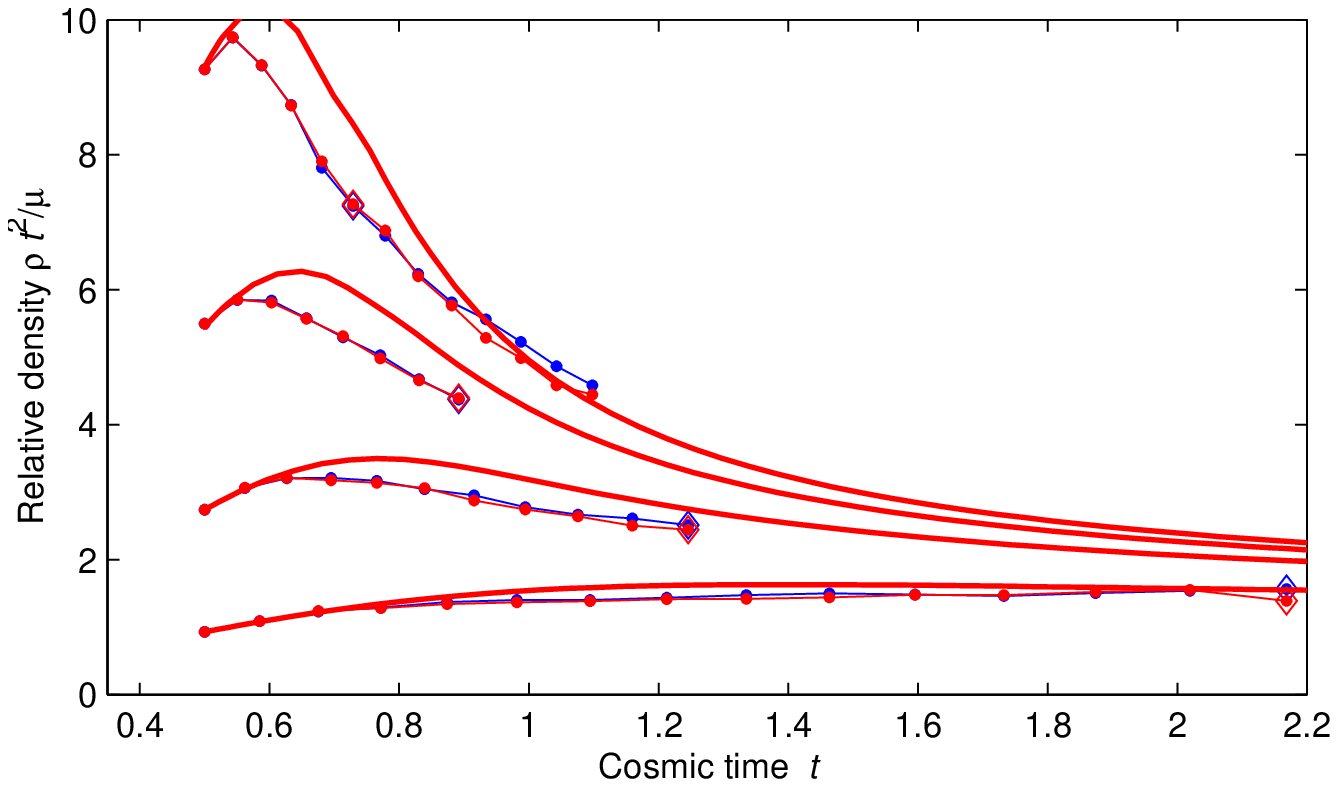}}
\centerline{
\epsfxsize=0.7\hsize\epsfbox{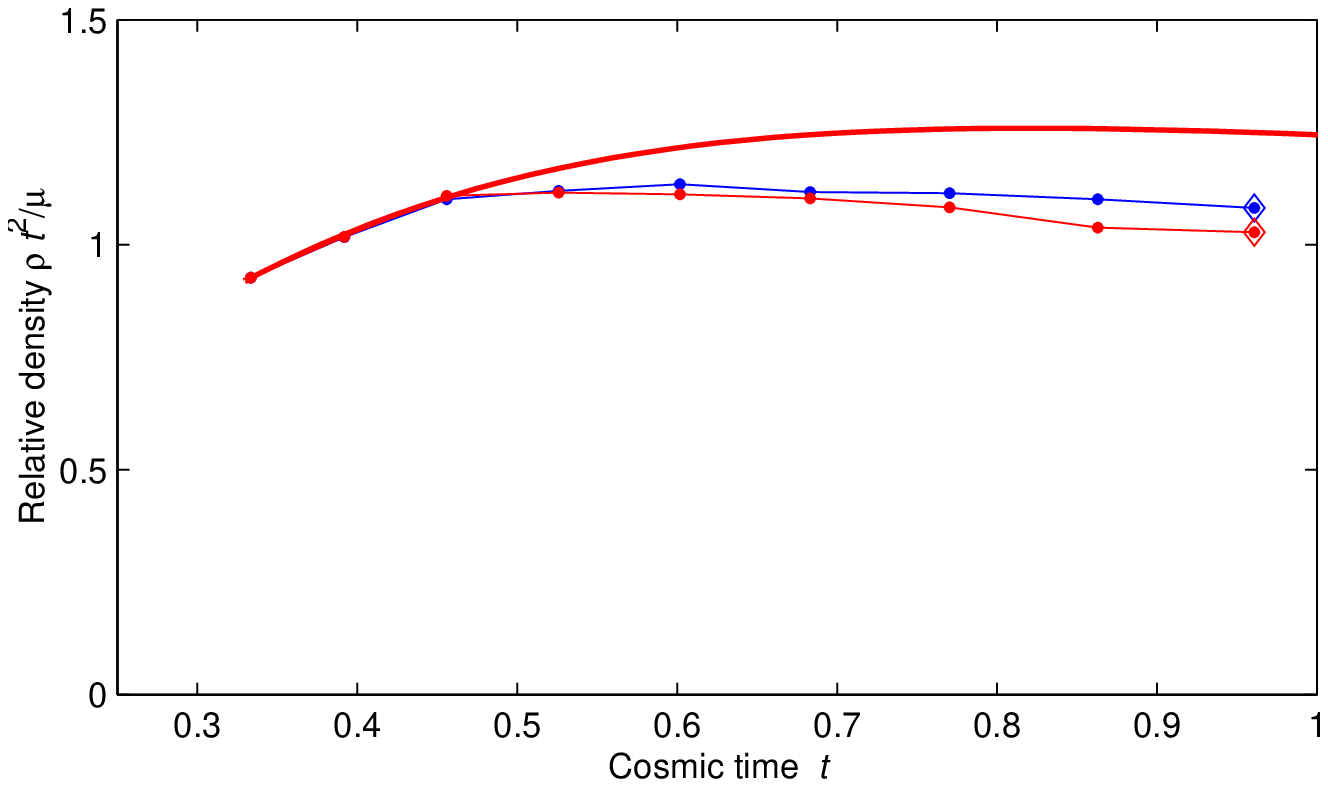}}
\vskip0in}
\caption{Analytic modelling of global strings in the radiation (top)
and matter (bottom) eras.
A good fit to the simulated data (red lines) was obtained with a VOS 
model (black lines) with the same loop production as local strings
(${\tilde c}=0.57$) and the addition of a large massless radiation damping 
coefficient ($\Sigma=1.1$).}
\label{global_vos}
\end{figure}

Finally, in Fig.~\ref{global_vos} we show an analogous plot for 
global string networks in the radiation  and matter eras. Here we retain 
the same loop chopping efficiency, ${\tilde c}=0.57$, but we
require a much larger radiation coefficient, namely $\Sigma\sim 1.1$,
to obtain satisfactory correspondence (in this case the radiation 
is massless with the radiation damping length 
$L_{\rm d}\rightarrow \infty$ in (\ref{grb2})). 
The fits to the string density 
are very good, but the deviations for the string velocities are 
slightly worse than for local strings. In passing however, we note
that if we had followed the velocity correction procedure outlined
above we could obtain an excellent fit for the global string
velocities (not just for the very early time evolution, but for
the asymptotic behaviour as well), albeit still with different
best-fit parameters.  This would correspond to incorporating the effect 
of the 
strong long-range forces in the VOS model 
which are clearly important for the small length-scales probed
by these simulations.

As expected, we can account for the large difference between the 
local and global string densities through the strong radiative damping
of the latter ($\Sigma \sim 1$).  The radiation back-reaction equivalent
of $\Gamma G\mu$ for the global string is the quantity \cite{Battye1994qa}
\begin{equation}
\kappa = \Gamma/2\pi \ln(L/\delta)\,,
\label{global_rad}
\end{equation}
where the average logarithmic cut-off for these simulations is certainly 
less than $\ln(L/\delta)\lesssim 3$.  Using the standard value for loops $\Gamma
\approx 65$ as an upper bound,
we have $\kappa \gtrsim 2$ which with $\Sigma\approx \kappa$
implies very strong radiative
damping for the simulated global strings (especially when compared with $\Gamma
G\mu \sim 10^{-4}$ for GUT-scale local strings).  We must be  
extremely careful, therefore, before extrapolating this small global/local string
density ratio from these simulations  to cosmological scales (for example,
as was assumed
in \cite{Yamaguchi1999yp,Yamaguchi1999dy}). The logarithmic term in
(\ref{global_rad}) for cosmological
global strings is much larger, typically 
with $\ln(L/\delta)\gtrsim 100$.  Thus the damping coefficient will fall
below $\Sigma \lesssim 0.1$, radiative effects will become perturbative
and loop production will dominate, so the VOS model predicts that 
cosmic local and 
global strings will actually have very comparable densities.  This point 
illustrates the value of a combined numerical and analytic approach
when attempting to describe 
these complex nonlinear systems in a cosmological context.

\subsection{Competing energy loss mechanisms}
\label{seccompeting}

\begin{figure}
\vbox{\centerline{
\epsfxsize=0.7\hsize\epsfbox{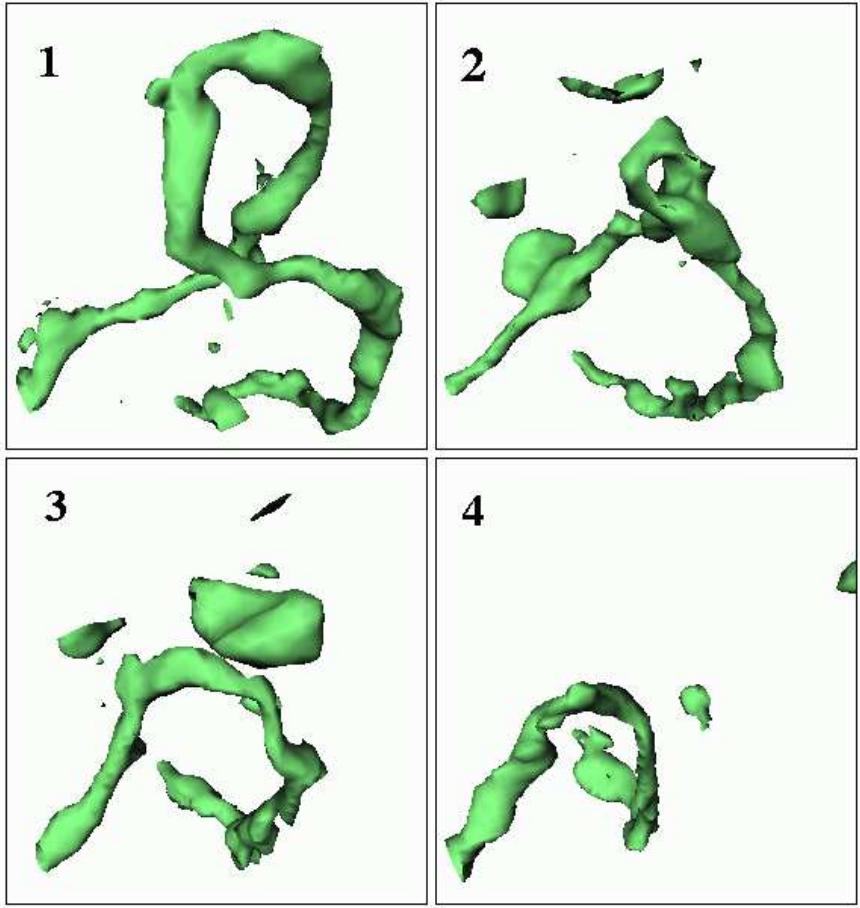}}
\vskip.4in}
\caption{Four stages of the formation of `proto-loops' in the field theory
simulations: A highly curved region of string (i), collapses to
form a nonlinear `lump' in the energy density (ii), which leaves the 
string (iii) to 
decay and disappear (iv).  This can only happen
 when the the string's topological 
winding inter-commutes
or annihilates in this region. Energy density contours are
plotted.}
\label{figprotoloop}
\end{figure}

The analytic modelling up to this point appears to be consistent
with the standard picture in which loop production is the primary 
decay mechanism.  It remains, however, to test this phenomenological
conclusion directly by measuring the actual energy loss into 
loops during the simulations.  In doing this, however, we face 
an obvious difficulty which requires a very different treatment 
of the loop distribution than in the usual `one scale' model.
If the standard cosmological picture of loop production
is correct, then we know from high resolution Nambu string
simulations that the typical loop creation scale $\bar\ell$ relative
to the horizon is $\alpha \equiv \bar\ell/t \lesssim 10^{-3}$ (if
gravitational radiation back-reaction sets this limit then $\alpha \sim 10^{-4}$
for GUT-scale strings).  The present field theory simulations,
however, have a dynamic range which is at least two orders of
magnitude poorer, so we cannot realistically hope to probe such
small-scale regimes. In consequence, we might be surprised to be
able to identify any loops at all in field theory simulations
and we certainly would not expect
their average creation size $\bar\ell$ to `scale' relative to $L$.
Instead, our numerical studies suggest that most loops are created with radii
comparable to the string thickness, with $\bar \ell \sim \pi$ almost
constant throughout. Simulation visualisations seem to indicate that
most energy is lost via such small loops or `proto-loops', that is, 
small-scale highly
nonlinear, but coherent, regions of energy density (as illustrated
in Fig.~\ref{figprotoloop}). Consistent with the Nambu simulations,
this `proto-loop' production occurs---like small loop formation---in
high curvature regions where the strings collapse and become very
convoluted \cite{Shellard1990evolcs}.

A proportion of these `proto-loops' have sufficient topology to be
identified as loops by our numerical diagnostics, so we can estimate
the energy loss via this pathway relative to other mechanisms.
Because these small loops are comparable to the string thickness, they
are strongly self-interacting and can be expected to be self-intersecting.
We can infer (and we see this in visual animations) that they 
annihilate very rapidly, decaying into 
massive particles almost as fast as allowed by causality.  Instead of
long-lived non-intersecting loops slowly decaying by gravitational 
radiation (with typical lifetimes $\tau \gtrsim 10^{-4}\bar\ell$), the 
simulation loops collapse and disappear like a circular loop in 
only half a period, that is,  in a time  $\tau \approx \ell/4$.  Hence, 
their time-dependent length,
can be described approximately by 
\begin{equation}
\ell \approx \ell_{\rm c} - 4(t-t_{\rm
  c})\,,\qquad 0< t-t_{\rm c} < \ell_{\rm c/4}
\end{equation}
where the loop
creation length  and time are $\ell_{\rm c}$ and $t_{\rm c}$. 
At any one time, then, the string energy loss through this rapid 
loop decay  can be simply
approximated by
\begin{equation}
\dot\rho_\infty \approx - 4\mu n_\ell\,,
\label{rapidloops}
\end{equation}
where
$n_\ell$ is the small loop number density.

For linear scaling to pertain in the `one-scale' model, a
dominant energy loss mechanism must behave as $\dot \rho_\infty\propto 
\rho_\infty/L \propto
t^{-3}$ (refer, for example,  to (\ref{rtl})).  Equating this 
with the overall small loop decay rate (\ref{rapidloops}), 
implies that we can achieve 
scaling with $n_\ell \propto t^{-3}$ or, equivalently, with
a loop energy density $\rho_\ell = \mu \bar \ell n_\ell \propto t^{-3}
\propto
\rho/t$ (with $\bar \ell$ constant).  Thus, with rapidly decaying
loops of small constant size, it is possible to maintain network scaling
while their relative contribution to the string energy density falls
dramatically as $\rho_\ell/\rho \propto t^{-1}$.  (Contrast this with 
the standard long-lived loop scenario in which the loop and infinite
string densities remain
proportional $\rho_\ell/\rho_\infty \propto {\rm const.}$)

\begin{figure}
\vbox{\centerline{
\epsfxsize=0.7\hsize\epsfbox{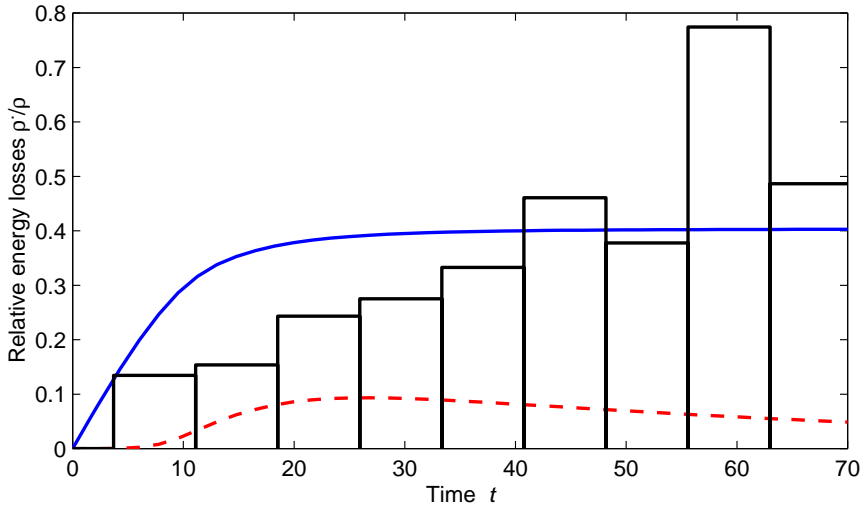}}
\vskip.4in}
\caption{The relative contributions of the different components
to the overall energy density losses of a flat space string network, as
predicted the analytic VOS model for the simulation in
Fig.~\ref{figflat_data} with $L_{\rm i} = 14.5$. The best fit 
parameters are $\tilde c =0.57$, $\Sigma=0.5$ and $L_{\rm d}= 4\pi$, 
and shown are loop creation losses (solid blue line) and massive 
radiation (dashed red line).
Superposed onto
this figure is a roughly normalised
 histogram of energy loss into loop production
estimated directly from measurements of the time-varying loop
density in this particular simulation.  It is apparent that the
loop energy loss contribution is becoming more important as the
simulation progresses.}
\label{figrelenergy}
\end{figure}

Fig.~\ref{figrelenergy} illustrates the relative small loop
contribution to the overall energy density losses throughout a
specific flat space simulation, that is, the case with $L_{\rm i} = 14.5$
shown in Fig.~\ref{figflat_data} (magenta line). This is
compared with our overall `best fit' VOS model in which 
loop creation is dominant but with some additional massive radiation.
 We can observe that the measured loop energy
losses grow steadily towards the analytic loop contribution, which
for these simulations we assume must include `proto-loops' both with
topology and without. We can see that the proportion of
loops---the topological `proto-loops'---gradually grows to meet or
even overtake the analytic loop contribution by the end of the
simulation. (Note that this correspondence is qualitative as we do
not have a precise measure of the typical small loop energy.)

We surmise that
loop/proto-loop production appears to provide a viable
explanation for the 
decay mechanism which maintains the scale-invariant evolution observed 
in the simulations.  However, this discussion is only a small initial step and 
needs to be made much more quantitative.  First, the simple analytic
model describing rapidly decaying loops needs to be developed further
and, secondly, loop production rates, distributions and decay timescales
need to be investigated in much greater detail in large field theory 
simulations.  Nevertheless, it is interesting to speculate about the 
clear qualitative trend that appears to be evident in Fig.~\ref{figrelenergy}. 
In extrapolating to cosmological scales we increase
the ratio of the correlation length to the string width by a further 
twenty orders of magnitude.  It is not unreasonable to expect, therefore, 
that the loop
contribution will become the completely dominant network decay mechanism.
 As the typical string
perturbation length scale grows and is affected by radiative
back-reaction, it is again reasonable to suppose that the typical
loop creation size will also grow, becoming many orders of magnitude
larger than the string thickness. Thus we could conclude that these field
theory simulations, once we account for their small dynamic range,
can be interpreted as being consistent with the standard picture 
of long string network
evolution via small loop production. Conversely, at the very least, the
simulations {\em do not} provide compelling evidence that
cosmological strings will decay primarily through the direct
radiation of ultra-massive particles.

\section{Conclusion}
\label{secconclusion}

We have explored and characterised scale-invariant string network
dynamics in numerical field theory simulations for local gauged
strings and global strings in flat space and, in an expanding background, 
in both 
radiation and matter eras.  We have estimated both the typical `scaling' energy 
density and the rms velocities of the strings for the dynamical ranges
currently accessible numerically.  
We have successfully modelled the
density and velocity of these string networks using a simple 
`one scale' analytic model (the VOS model), describing network
decay into small loops, supplemented with some direct massive radiation. 
We conclude, from these analytic fits and a qualitative examination
of loop production, that these results produce an apparently coherent
and adequate model for string network evolution.
However, this is not to suggest that
we have provided a complete or detailed description of the complex
nonlinear processes that underlie network evolution on these small
scales and many avenues of research remain to be investigated.

An interesting outstanding issue is the close correspondence between 
the loop creation efficiency $\tilde c$ in flat space Nambu simulations
and abelian-Higgs simulations in both flat and expanding backgrounds.
Field theory simulations do not appear to have sufficient resolution at 
present to allow for the accumulation of the small-scale structure so 
evident in expanding universe Nambu simulations.  The `fat string'
algorithm we have presented can be used to extend the dynamic range 
of simulations, but the smallest physical scales in this case also grow.
Mesh refinement techniques are likely to be necessary in order to 
make major advances over the present work.

These results have significant cosmological implications. Our
expectation is that massive particles will only be produced
infrequently in highly nonlinear string regions, such as at cusps
and re-connections. The ensuing flux of cosmic rays should be
relatively low \cite{Bhattacharjee1990js,Bhattacharjee1998qc,Protheroe1996pd,Berezinsky1998qv}.
It now appears that some estimates of the ensuing cosmic ray flux from direct
massive radiation from strings were overly
optimistic \cite{Vincent1998cx}.
This work points to the need for caution in making cosmological
extrapolations from small-scale numerical simulations and to the
need for further progress understanding string radiation
back-reaction.

\section*{Acknowledgements}
We are grateful for useful discussions with Brandon Carter, Mark
Hindmarsh, Graham Vincent, Richard Battye and Jose Blanco-Pillado. 
C.\ M.\ is funded by FCT (Portugal), grant no.\ FMRH/BPD/1600/2000.
The simulations were performed on the
COSMOS Origin2000 supercomputer which is supported by Silicon
Graphics, HEFCE and PPARC.

\end{document}